\documentclass[11pt]{article}
\pdfoutput=1

\usepackage{stolenstyle}

\usepackage{amsmath,epsf,amssymb,latexsym,amsthm,setspace,bbm,array,pifont,enumerate}

\DeclareFontFamily{U}{rsf}{}
\DeclareFontShape{U}{rsf}{m}{n}{
  <5> <6> rsfs5 <7> <8> <9> rsfs7 <10-> rsfs10}{}
\DeclareMathAlphabet\Scr{U}{rsf}{m}{n}

\def\AC#1#2{{\{#1,#2\}}}

\def\cDb{{\overline{\cD}}}
\def\cQb{{\overline{\cQ}}}

\def\GUL{\GU(1)_{\text{L}}}
\def\GUR{\GU(1)_{\text{R}}}

\def\C{{\mathbb C}}

\def\R{{\mathbb R}}
\def\Z{{\mathbb Z}}

\def\Pic{\operatorname{Pic}}

\def\GU{\operatorname{U{}}}

\def\p{\partial}
\def\pb{\bar{\partial}}

\def\ff#1#2{{\textstyle\frac{#1}{#2}}}

\def\cA{{\cal A}}

\def\cD{{\cal D}}
\def\cE{{\cal E}}
\def\cF{{\cal F}}
\def\cG{{\cal G}}
\def\cH{{\cal H}}

\def\cK{{\cal K}}
\def\cL{{\cal L}}
\def\cM{{\cal M}}

\def\cO{{\cal O}}

\def\cQ{{\cal Q}}
\def\cR{{\cal R}}

\def\cU{{\cal U}}

\def\cW{{\cal W}}

\def\cY{{\cal Y}}

\newcommand\alphab{\overline{\alpha}}
\newcommand\betab{\overline{\beta}}
\newcommand\gammab{\overline{\gamma}}

\newcommand\thetab{\overline{\theta}}

\newcommand\phib{\overline{\phi}}

\newcommand\etat{\widetilde{\eta}}

\newcommand\mut{\widetilde{\mu}}

\newcommand\Gammab{\overline{\Gamma}}

\newcommand\Phib{\overline{\Phi}}

\newcommand\Psit{\widetilde{\Psi}}

\newcommand\ib{\overline{\imath}}
\newcommand\jb{\overline{\jmath}}
\newcommand\kb{\overline{k}}

\newcommand\ub{\overline{u}}

\newcommand\yb{\overline{y}}
\newcommand\zb{\overline{z}}

\newcommand\Ab{\overline{A}}

\newcommand\Lb{\overline{L}}

\newcommand\Tb{\overline{T}}
\newcommand\Ub{\overline{U}}

\def\bq{{\textbf{q}}}

\def\cYb{{{\overline{\cY}}}}

\def\cWb{{{\overline{\cW}}}}

\def\bY{{\boldsymbol{Y}}}
\def\bY{{Y}}

\theoremstyle{definition}

\usepackage{tikz}

\usetikzlibrary[shapes.geometric]
\usetikzlibrary{positioning} 
\usetikzlibrary{calc,intersections,through,backgrounds}
\usetikzlibrary{cd}

\tikzset{>=stealth}
\tikzset{every picture/.style={very thick}}

\def\cWb{{\overline{\cW}}}

\def\bq{{\textbf{q}}}

\def\sleft{\text{\tiny{L}}}
\def\sright{\text{\tiny{R}}}

\def\preprint{ ~~ }

\title{Marginal deformations of Calabi-Yau hypersurface hybrids with (2,2) supersymmetry}
\author[a] {Griffen Adams}
\author[b] {and Ilarion V.~Melnikov}
\affiliation[a] {Department of Physics, University of Connecticut, Storrs, CT 06269-3046, USA}
\affiliation[b] {Department of Physics and Astronomy,
James Madison University,
Harrisonburg, VA 22807, USA}

\emailAdd{griffen.adams@uconn.edu}
\emailAdd{melnikix@jmu.edu}

\abstract{We study two-dimensional non-linear sigma models with (2,2) supersymmetry and a holomorphic superpotential that are believed to flow to unitary compact (2,2) superconformal theories with central charges $c_{\sleft} = c_{\sright} = 9$.  The SCFTs have a set of marginal deformations, and some of these can be realized as deformations of parameters of the UV theory, making it possible to apply techniques such as localization to probe the deformations of the SCFT in terms of a UV Lagrangian.  In this work we describe the UV lifts of the remaining SCFT infinitesimal deformations, the so-called non-toric and non-polynomial deformations.  Our UV theories naturally arise as geometric phases of gauged linear sigma models, and it may be possible to extend our results to find lifts of all SCFT deformations to the gauged linear sigma model.}

\begin{document}

\maketitle

\section{Introduction} \label{s:Introduction}

Calabi-Yau compactification plays a central role in the study of string theory in all of its guises and duality frames.  One of the oldest of these is the realization that compact Calabi-Yau manifolds have an intimate relation to non-trivial two-dimensional superconformal theories (SCFTs) with (2,2) supersymmetry.  Indeed, there is a well-motivated conjecture that given a compact smooth Calabi-Yau manifold $X$, the (2,2) supersymmetric non-linear sigma model with target space $X$ can be endowed with a smooth K\"ahler metric $g$ and a closed Kalb-Ramond field $B$ such that the resulting theory is superconformal.  More precisely, it is believed that for a fixed choice of complex structure and complexified K\"ahler class  there is a unique K\"ahler metric $g$ compatible with these structures such that the non-linear sigma model with target space $X$ and metric $g$ is a superconformal field theory~\cite{Nemeschansky:1986yx}.  Furthermore, when the volume of $X$ is taken to be large in string units the metric on $g$ approaches the unique Ricci-flat Calabi-Yau metric for the fixed complex structure and K\"ahler class.  These familiar notions are of course textbook material~\cite{Polchinski:1998rr} discussed in detail in many classic reviews such as~\cite{Greene:1997my,Hori:2003ds}.

Except at special points in the moduli space, the SCFTs obtained in this way are not solvable, and this is both the challenge and the appeal of the construction.  To probe the physics of the putative SCFT a number of powerful methods have been devised.  Many of these approaches are unified via (2,2) gauged linear sigma models introduced in this context in~\cite{Witten:1993yc}: for an appropriate choice of parameters these two-dimensional gauge theories are believed to flow to the same SCFT as given by the non-linear sigma model with target space $X$, where $X$ is realized as a complete intersection in a toric (or, in the non-abelian case, a Grassmannian or closely related) variety.  Techniques based on topological field theory and localization can then be used to probe the strongly-coupled IR dynamics described by the SCFT via UV computations based on a Lagrangian gauge theory.  A recent review of these constructions was given in~\cite{Melnikov:2019tpl}.  

The results based on such computations have had a profound impact on both mathematical physics, primarily through applications to mirror symmetry, as well as on string compactification, but all such applications have an important caveat:  they can only be applied to those IR computations that have a simple UV lift.  In this work we tackle the simplest but perhaps also the foundational aspect of this problem: typically a UV lift does not describe the full deformation space of the putative superconformal theory.  The issue was already clear in the earliest large-scale constructions of such Calabi-Yau manifolds, the so-called ``CICY'' manifolds obtained as complete intersections in products of projective spaces~\cite{Candelas:1987kf,Green:1987rw,Hubsch:1992nu}: in general when $X$ is obtained as such as space, only a subset of deformations of complex structure is obtained from deformations of the defining polynomial equations.  This subset is exactly the set that has a good presentation in a corresponding linear sigma model, and we are then faced with a general question:  can we describe the remaining deformations in terms of the fields of the UV theory?

We will describe a solution to this problem relevant to another large class of Calabi-Yau compactifications:  hypersurfaces in toric varieties.  These manifolds were introduced by Batyrev in the context of mirror symmetry~\cite{Batyrev:1994hm} and subsequently generalized to complete intersections in toric varieties~\cite{Batyrev:1994pg}.  These models have a simple gauged linear sigma model presentation~\cite{Morrison:1994fr}, and their moduli spaces give a precise distinction between those moduli with a simple UV lift, and those that do not have such a lift.  Explicitly, the moduli space of the SCFT is locally a product of two special K\"ahler manifolds $\cM^{\text{ac}} \times \cM^{\text{cc}}$, with the first factor describing the complexified K\"ahler deformations associated to the (a,c) ring of the SCFT, while the second factor describes the complex structure deformations associated to the (c,c) ring of the SCFT.  There is then a canonical identification between the tangent spaces to these with cohomology groups on $X$:
\begin{align}
T_{\cM^{\text{ac}}} &\simeq H^1(X,\Omega_X^1)~,&
T_{\cM^{\text{cc}}} & \simeq H^1(X,T_X)~.
\end{align}
Here $T_X$ is the (holomorphic) tangent sheaf on $X$, while $\Omega_X^1$ is its dual, which can also be thought of as the   sheaf of (1,0)-forms on $X$.  The dimensions of the spaces are then given by
\begin{align}
\dim \cM^{\text{ac}}  &= h^1(X,\Omega_X^1) = h^{1,1}(X)~,&
\dim \cM^{\text{ac}}  &= h^1(X,T_X) = h^{1,2}(X)~.
\end{align}
Furthermore, there is a decomposition~\cite{Aspinwall:1993rj,Batyrev:1994hm}, nicely reviewed in~\cite{Cox:2000vi}, 
\begin{align}
h^{1,1}(X) & = h^{1,1}_{\text{toric}}(X) + h^{1,1}_{\text{non-toric}}(X)~,&
h^{1,2}(X) & = h^{1,2}_{\text{poly}}(X) + h^{1,2}_{\text{non-poly}}(X)~.
\end{align}
The first term in each equation has a straightforward interpretation in the gauged linear sigma model:  each ``toric'' deformation can be understood as a deformation of the complexified Fayet-Iliopoulos parameters encoded in the twisted chiral superpotential, while each polynomial deformation can be understood as a deformation of the chiral superpotential determined by the defining equation of the hypersurface.  In (2,2) SCFTs the decompositions turn out to be mirror-symmetric, i.e. mirror symmetry exchanges the toric deformations of the original theory with the polynomial deformations of the mirror.\footnote{This is not preserved by (0,2) marginal supersymmetric deformations~\cite{Kreuzer:2010ph,Melnikov:2010sa}.}  On the other hand, the remaining deformations do not have a simple UV description.

There are several ways to address this issue.  The most pragmatic is simply to stick to examples where the non-toric and non-polynomial deformations are absent.  This is for example done in the CICY literature by restricting to what are termed ``favorable'' configurations.  On the other hand, if one's interest is in a particular $X$, then one may try to find a more general construction that presents $X$ as a complete intersection in some other variety where the analogues of the non-toric and non-polynomial deformations are absent.  An early approach of this sort was made in~\cite{Berglund:1995gd}; the more recent construction of ``generalized CICYs''~\cite{Anderson:2015iia} offers another set of promising candidates for finding such generalizations.  However in general it is not obvious how to construct such a desired UV theory given a particular $X$.  

In addition, there is an important question of principle.  Given an RG flow from a (2,2) UV theory to the SCFT we can ask whether it is possible to describe marginal operators in the SCFT in terms of operators constructed from the UV fields based on a classical Lagrangian.  We know that in general this is too much to ask: for example, a classical field theorist equipped with the Lagrangian of a compact boson will be hard-pressed to discover the marginal operators that exist at the self-dual radius!  However, we can hope that the situation is under better control when the flow leads to a weakly-coupled large radius non-linear sigma model with target space $X$, and we will see our hope borne out, so that we will be able to present candidate operators in the chiral algebra of a UV theory that describe the full set of marginal deformations of the IR SCFT.

We term the UV theory we study a ``hypersurface hybrid theory.''\footnote{In some literature---for example~\cite{Guffin:2008kt,Jarvis:2014,Li:2019}---these theories are referred to as Landau-Ginzburg models.  We choose not to use that terminology to emphasize the significant differences between a Landau-Ginzburg theory and a curved non-linear sigma model.}  Such a theory arises as a phase in the gauged linear sigma model for a hypersurface $X$ in a toric variety $V$, and it can be formally obtained by taking the linear sigma model deep in a geometric phase and sending the gauge coupling to infinity while keeping the chiral superpotential couplings finite.  The result is a non-linear sigma model with target space $\bY$, the total space of the canonical bundle $\cO_V(K_V) \to V$ equipped with a chiral superpotential $\cW = \Phi P$, where $\Phi$ is the distinguished fiber coordinate, and $P$ is a section of the anticanonical bundle $\cO_V(-K_V) \to V$ chosen to be suitably generic so that $X = \{P = 0\} \subset V$ is a smooth manifold.  We can think of this theory as an example of a hybrid theory, such as those introduced in~\cite{Bertolini:2013xga} and recently studied in a number of works including~\cite{Erkinger:2022sqs}.  A generic hybrid theory is constructed as a fibration of a (2,2) Landau-Ginzburg theory over a suitable base manifold $V$, where the Landau-Ginzburg fields are sections of certain vector bundles over $V$, while the Landau-Ginzburg superpotential varies holomorphically over the base.  Our theory is a special and rather degenerate case, where the Landau-Ginzburg potential is linear in the fiber field.  This means, for example, that we cannot think of the theory as a fibration of a \textit{supersymmetric} Landau-Ginzburg model on a curved base.  A related property is that the introduction of the superpotential term decreases the value of the central charge, while in the hybrid theories considered in~\cite{Bertolini:2013xga} the Landau-Ginzburg degrees of freedom made a non-negative contribution to the central charge.  Nevertheless, we will see that much of the technology developed in~\cite{Bertolini:2013xga} continues to apply in this case and gives a computable framework, in particular for the Ramond sector of the theory.

Our central result is to find explicit representatives for non-toric and non-polynomial deformations of the SCFT in the chiral algebra of the theory:  the cohomology of the right-moving supercharge, or, in a superfield formulation, the cohomology of the super-covariant derivative $\cDb$, which we denote by $\cH_{\cDb}$.  Working in the classical UV theory we obtain subspaces $\cH_{\cDb}^{\text{ac}}$ and $\cH_{\cDb}^{\text{cc}}$ which we expect to flow to marginal (a,c) and (c,c) operators in the SCFT.  While already solving a problem in principle, we primarily view this result as a positive step in providing a similar description of deformations at the level of the gauged linear sigma model.

The rest of the note is organized as follows: we introduce some (2,2) superspace notation in section~\ref{s:CYwarmup} and then apply it to give a large radius description of deformations in a compact Calabi-Yau non-linear sigma model.  Next we give a discussion of deformations of a hypersurface $X$ in $V$ in terms of algebraic geometry and phrase the non-toric and non-polynomial deformations solely in terms of properties of algebraic geometry of $V$.  Our key results are then obtained in section~\ref{s:Hybriddefs}, where we lift these deformations to the hypersurface hybrid based on the non-linear sigma model with target space $\bY$.  In section~\ref{s:HybridNSR} we re-examine the marginal deformations by studying the NS-R sector of the theory via the techniques of~\cite{Bertolini:2013xga} and reproduce the results obtained in previous sections.  We conclude with a discussion of future directions.

\acknowledgments   IVM's work is supported in part by the Humboldt Research Award 
as well as the Educational Leave program at James Madison University.  Our work on this project was also supported by the NSF Grant PHY-1914505.
We thank  P.~Aspinwall and R.~Plesser for useful discussions.  IVM acknowledges an ancient collaboration with B.~Wurm that attempted to tackle some closely related questions.

\section{Warm up: deformations of Calabi-Yau non-linear sigma models} \label{s:CYwarmup}
In this section we set out the notation that we will use in the rest of our note, and we will illustrate the basic ideas in the familiar setting of the chiral algebra of a large radius compact Calabi-Yau manifold.  Additional details may be found in, for example,~\cite{Bertolini:2013xga,Melnikov:2019tpl}.

\subsection{Superspace conventions}
Our conventions for superspace are those of~\cite{Melnikov:2019tpl}.  We work in Euclidean signature on a flat worldsheet $\Sigma = \C$ and (2,2) superspace coordinates $(z, \theta',\thetab'; \zb, \theta,\thetab)$.  Using the short-hand notation $\p_z = \frac{\p}{\p z}$ and $\pb_{\zb} = \frac{\p}{\p\zb}$, a representation of the right-moving (or anti-holomorphic) supersymmetry algebra is furnished by the antiholomorphic superspace derivatives and supercharge operators
\begin{align}
\label{eq:EucsuperderR}
\cD & = \p_{\theta} + \thetab \pb_{\zb}~, 
&
\cQ & = \p_{\theta} - \thetab \pb_{\zb}~, \nonumber\\[2mm]
\cDb & = \p_{\thetab} + \theta \pb_{\zb}~, 
&
\cQb & =  \p_{\thetab} - \theta \pb_{\zb}~.
\end{align}
The non-trivial anti-commutators for these are 
$\AC{\cD}{\cDb} = 2\pb_{\zb}$ and $\AC{\cQ}{\cQb} = -2\pb_{\zb}$. We also have their ``holomorphic'' versions
\begin{align}
\label{eq:EucsuperderL}
\cD' & = \p_{\theta'} + \thetab' \p_z~, 
&
\cQ' & = \p_{\theta'} - \thetab' \p_z~, \nonumber\\[2mm]
\cDb' & = \p_{\thetab'} + \theta' \p_z~, 
&
\cQb' & =  \p_{\thetab'} - \theta' \p_z~,
\end{align}
which have non-trivial anti-commutators $\AC{\cD'}{\cDb'} = 2\p_z$ and $\AC{\cQ'}{\cQb'} = -2\p_z$. 

We will be working with Lagrangian field theories based on bosonic chiral superfields $\cY^\alpha$, which satisfy the constraints
\begin{align}
\cDb \cY^\alpha & = 0~,&
\cDb' \cY^\alpha & = 0~,
\end{align}
as well as their anti-chiral charge-conjugates $\cYb^{\alphab}$, which are annihilated by $\cD$ and $\cD'$.  These fields have superspace expansions
\begin{align}
\cY^\alpha &= y^\alpha + \cdots~,&
\cYb^{\alphab} & = \yb^{\alphab}+\cdots~,
\end{align}
and the bosonic fields $y^\alpha(z,\zb)$ and $\yb^{\alphab}(z,\zb)$ take values in a K\"ahler manifold $\bY$---the target space of the non-linear sigma model.  Denoting the embedding map as $f: \Sigma \to \bY$, the  superfields  $\cD \cY^\alpha$ and $\cDb \cYb^{\alphab}$ have spin $-1/2$ and take values in pullback bundles $f^\ast (T_{\bY})$ and $f^\ast(\Tb_{\bY})$ respectively.\footnote{ More precisely, these superfields take values in appropriate pullbacks of the target space tangent bundle $T_{\bY}$ tensored with a spin bundle on the worldsheet.  For example, $\cD \cY^\alpha$ is a section of $f^\ast(T_{\bY}) \otimes \overline {K}^{1/2}_{\Sigma}$. In our case the canonical bundle $K_\Sigma$ and its conjugate $\overline{K}_\Sigma$ are trivial, and so are the spin bundles.  The spin bundles play an important role when we place the theory on a curved worldsheet, when we consider topologically non-trivial field configurations, or when we perform a topological twist to obtain a cohomological topological field theory see, e.g.~\cite{Deligne:1999qp,Hori:2003ds}.  These subtleties will not play a role in our classical considerations, and we will just keep track of the spin eigenvalues.}  In what follows we will lighten notation and not write the explicit pullback by the map $f$.  The superfields $\cD' \cY^\alpha$ and $\cDb'\cYb^{\alphab}$ are valued in the same bundles as their anti-holomorphic counter-parts but carry spin $+1/2$.  This geometric structure allows us to formulate the action and path integral of the non-linear sigma model.  Given a cover $\{ \mathfrak{U}_a\}_{a\in I}$ of $\bY$ with holomorphic transition functions relating the coordinates in overlapping patches $\mathfrak{U}_a\cap \mathfrak{U}_b\neq \emptyset$ as $y_a^\alpha = F^\alpha_{ab}(y_b)$, the superfields of the theory transform accordingly:
\begin{align}
\cY^\alpha_a & = F^\alpha_{ab}(\cY_b)~,&
\cYb^{\alphab}_a & = \overline{F^\alpha_{ab}(\cY_b)}~,
\end{align}
and the superspace derivatives transform covariantly.  For example,
\begin{align}
(\cD \cY^\alpha)_a & = \frac{\p F^\alpha_{ab}}{\p \cY_b^\beta} (\cD\cY^\beta)_b~.
\end{align}
Note that this implies that the higher order terms in the $\theta,\theta'$ expansion of the superfields pick up connection terms in their transformations.  This is a familiar feature of superspace~\cite{West:1990tg,Wess:1992cp}.  

Before we proceed further we will fix some notation for a holomorphic vector bundle $\cE$ over a K\"ahler manifold $Y$ of dimension $d$. We denote the dual bundle by $\cE^\ast$, and the complex conjugate bundle by $\overline{\cE}$.   We denote by $\cA_Y^{p,q}(\cE)$ the vector space of sections of (p,q) forms on $Y$ valued in $\cE$.   The vector space of (p,q) forms on $Y$ will just be denoted by $\cA_Y^{p,q}$.  Although we will primarily work in the smooth category, it will be useful for us to think of the bundles as sheaves.  We will denote by $\cO_Y$ the structure sheaf on $Y$, $T_Y$ will be the tangent sheaf, and $\Omega^p_Y$ the sheaf of $(p,0)$-forms; of course $\Omega^1_Y = T_Y^\ast$.  $K_Y$ denotes the canonical divisor on $Y$.  For all examples we consider $K_Y$ will be Cartier, so that $\cO_Y(K_Y) \simeq \Omega^{d}_Y$ is the canonical bundle.  When $Y$ is compact and smooth we will frequently make use of the isomorphism between Dolbeault and \v{C}ech cohomology groups $H^{p,q}_{\pb}(Y,\cE) \simeq H^q(Y,\Omega_Y^p \otimes \cE)$.  The former naturally show up in the physical theory, while the latter are computationally more accessible.

\subsection{The action and its key symmetries}
To write down an explicit action, we fix a K\"ahler metric $\cG$ on $\bY$, locally given by a K\"ahler potential $\cK$.  It is also possible to include the coupling to a closed $B$-field, but this will not play a role in our classical discussion.  If $\bY$ is non-compact and admits a global holomorphic function $\cW(\cY)$, then we can also add a chiral superpotential term to the action.  Letting $m$ be a parameter with units of mass, the standard $2$-derivative action with a chiral superpotential is
\begin{align}
S = S_{\text{kin}} + S_{\text{pot}}~,
\end{align}
with
\begin{align}
\label{eq:NLSMAction}
S_{\text{kin}} & = \ff{1}{4\pi} \int d^2z \underbrace{ \cD \cDb \cDb' \cD'}_{= \cD_{\text{tot}}}
\left[ \ff{1}{2} \cK (\cY ,\cYb) \right]~, &
S_{\text{pot}} & = \ff{m}{4\pi} \int d^2z\, \cD \cD' \cW(\cY) + \text{h.c}~.
\end{align}
It is understood that the Grassmann coordinates $\theta,\theta'$ and their conjugates are to be set to zero after all of the superspace derivatives are taken.  

While the kinetic term is defined only patch by patch, the action is nevertheless well-defined since K\"ahler transformations give rise to terms annihilated by the superspace derivatives.  Moreover, because the fermion couplings are non-chiral the action and the path integral are invariant under (complex) diffeomorphisms, which allows both to be well-defined despite the curvature of the target space.  As we will be mostly concerned with aspects of the classical field theory, the equations motion will play an important role.  In superspace these take the form
\begin{align}
\label{eq:EOM}
\cDb' \cDb \cYb^{\betab} & = - \Gammab^{\betab}_{\alphab\gammab} \cDb' \cYb^{\alphab} \cDb \cYb^{\gammab} + 2 m \cG^{\betab \alpha} \cW_\alpha~,
\nonumber\\[2mm]
\cD' \cD \cY^\alpha & = -\Gamma^{\alpha}_{\beta\gamma} \cD' \cY^\beta \cD \cY^\gamma
-2 m \cWb_{\alphab} \cG^{\alphab \alpha} ~.
\end{align}
Here $\cG^{\betab\alpha}$ denotes the inverse K\"ahler metric, while $\Gamma$ and $\Gammab$ are the Chern connections on $T_{\bY}$ and $\Tb_{\bY}$ respectively:
\begin{align}
\Gamma^{\alpha}_{\beta\gamma} &= \cG^{\alphab \alpha} \cG_{\beta \alphab, \gamma}~, &
\Gammab^{\alphab}_{\betab\gammab} & = \cG^{\alphab \alpha} \cG_{\alpha\betab,\gammab}~.
\end{align}
The notation $\cW_\alpha$ is a short-hand for the components of $\p \cW =  \frac{\p W}{\p y^\alpha} d y^\alpha$, and similarly $\cWb_{\alphab}$ denote the components of $\pb \cWb$.\footnote{ Our notation for the spacetime Dolbeault differential operators $\p$ and $\pb$, with $\text{d} = \p + \pb$ is close to the world-sheet derivatives $\p_z$ and $\pb_{\zb}$; we hope the subscripts on the latter will lessen the confusion.}  We will find another form of the equations of motion useful as well:
\begin{align}
\label{eq:EOMK}
\cDb\,\cDb'  \cK_\alpha & = -2m \cW_\alpha~,&
\cD\cD'  \cK_{\alphab} & = 2 m \cWb_{\alphab}~.
\end{align}
Here again $\cK_\alpha = \frac{\p \cK}{\p y^\alpha}$, and $\cK_{\alphab} = \frac{\p \cK}{\p \yb^{\alphab}}$.  In this notation the K\"ahler metric is $\cG_{\alpha\betab} = \cK_{\alpha\betab}$.

When $\cW = 0$ the action has a classical $\GU(1)_{\sleft}\times\GU(1)_{\sright}$ global R-symmetry with the following action:
\begin{align}
\begin{matrix}
~			&  \theta' 	& \thetab' 	& \theta 	&\thetab 		& \cY^\alpha \\[2mm]
\GU(1)_{\sleft}	& +1		&-1		& ~0		&~0			& ~0 \\[3mm]
\GU(1)_{\sright}	& ~0		& ~0		& +1 	&-1			& ~0
\end{matrix}
\end{align}
These symmetries are chiral and in general anomalous, and the anomaly is proportional to $c_1(T_{\bY})$.  However, we will insist that the canonical bundle of $\bY$ is trivial, i.e. $\cO_{\bY}(K_{\bY}) = \cO_{\bY}$, which implies $c_1(T_{\bY}) = 0$.  While the superpotential in general breaks the symmetry, if $\bY$ and $\cK$ admit a holomorphic Killing vector $v$ such that the Lie derivative with respect to $v$ preserves the superpotential, i.e. $\cL_{v} \cW = \cW$, then the action will preserve a modified $\GU(1)_{\sleft}\times\GU(1)_{\sright}$ symmetry, which has the infinitesimal action
\begin{align}
\delta \theta' & = i \alpha_{\sleft} \theta'~,&
\delta \theta & = i\alpha_{\sright}\theta~,&
\delta \cY^\alpha & = i\alpha_{\sleft} \cL_{v} \cY^\alpha + i \alpha_{\sright} \cL_{v} \cY^\alpha~.
\end{align}
In all of our theories the symmetries will be compact and turn out to act with integral charges on a natural basis of the fields.
These symmetries will be crucial in making the connection between the UV theory and the IR SCFT.  In particular, we will assume that these symmetries flow to  the $\GU(1)_{\sleft} \times\GU(1)_{\sright}$ R-symmetries of the (2,2) superconformal algebra.  This seems to be a sound assumption for flows to compact and unitary (2,2) SCFTs, made implicitly or explicitly in most Lagrangian constructions of such theories.  In the more general case of (0,2) supersymmetric flows the circumstances when this assumption is justified remain to be understood~\cite{Bertolini:2021hal}.

\subsection{A view of Calabi-Yau deformations}
In this section we develop some of the main tools that we will use in our study of marginal deformations in a particularly well-understood context:  the deformations of the SCFT associated to a smooth compact Calabi-Yau manifold $\bY$ based on a large-radius non-linear sigma model description.  Our perspective is particularly inspired by~\cite{Beasley:2004ys}, as well as observations on conformal perturbation theory such as those given in~\cite{Green:2010da} in a four-dimensional context.

Before proceeding, we fix our definition of a compact Calabi-Yau manifold as a smooth K\"ahler manifold $\bY$, $\dim_\C Y =d$, with trivial canonical bundle and $H^i(\bY,\cO_{\bY}) = 0$ for $0<i < d-1$.  The last condition excludes cases such as $T^6$ or $\text{K3}\times T^2$~: there is a good physical reason to do this, since in those cases the superconformal algebra is enhanced, which leads to a different structure of the moduli space of marginal deformations.  A recent discussion of the SCFT moduli space in this enhanced context can be found in~\cite{Gomis:2016sab}.

With these definitions fixed, we consider the perspective of conformal perturbation theory:  the marginal deformations of a (2,2) superconformal theory are encoded in the deformation of an ``action''\footnote{This does not require a Lagrangian definition of the SCFT: more generally $\Delta S$ is used to define (after suitable regularization and renormalization) the perturbed correlation functions via $\langle \cdots  e^{-\Delta S} \rangle$.}
\begin{align}
\Delta S_{\text{CFT}} & =  \int d^2 z\, \cD\cDb' \Psit(z,\zb) +  \int d^2z\, \cD \cD' \Psi~+\text{h.c.},
\end{align}
where $\Psit$ is an (a,c) field with $\GU(1)_{\sleft}\times\GU(1)_{\sright}$ charges $(-1,1)$, while $\Psi$ is a (c,c) field with $\GU(1)_{\sleft}\times\GU(1)_{\sright}$ charges $(1,1)$.

When working in a large radius limit, meaning the typical length scale of the compactification geometry is much larger than the string length, we should be able to use the non-linear sigma model fields to describe the spectrum of (a,c) and (c,c) operators and the associated (infinitesimal) deformations, and we will reproduce the familiar results relating the space of infinitesimal deformations to Dolbeault cohomology on $\bY$:
\begin{align}
T_{\cM^{\text{cc}}} &\simeq H^{0,1}_{\pb} (\bY, T_{\bY})~, &
T_{\cM^{\text{ac}}} &\simeq H^{0,1}_{\pb} (\bY, T^\ast_{\bY})~. &
\end{align}
We begin with the (a,c) deformations.  The superfield $\Psit$ should have the following properties:
\begin{enumerate}
\item $\Psit$ is well-defined on the NLSM target space and is expressed in terms of the superfields $\cY,\cYb$, and their superspace derivatives;
\item it has $\GUL\times\GUR$ charges $(-1,1)$;
\item it carries (classical) dimensions $(h_{\sleft},h_{\sright}) = (\ff{1}{2},\ff{1}{2})$;
\item $\Psit$ is twisted chiral up to the NLSM equations of motion.
\end{enumerate}
Denoting the space of $(p,q)$ forms on $\bY$ by $\cA^{p,q}_\bY$, we find that properties 1,2,3 imply
\begin{align}
\Psit & = \omega_{\alpha\betab} \cD' \cY^\alpha \cDb \cYb^{\betab} + \cD' \cDb f~,
\end{align}
where $\omega \in \cA^{1,1}_\bY$, and $f \in \cA^{0,0}_\bY$.  Property 4 holds if and only if $d \omega = 0$. 

Before we continue, we point out a frequent super-abuse of notation.  We will often discuss a geometric quantity, for example the form
\begin{align}
\omega = \omega_{\alpha\betab} (y,\yb) dy^\alpha\wedge d\yb^{\betab}~,
\end{align}
that we will use to construct a superfield expression such as $\Psit$.  In the latter it should be understood that we replace the coordinate dependence by the corresponding superfields, so that we should really write (already omitting the pullback to the worldsheet!)
\begin{align}
\Psit & = \omega_{\alpha\betab}(\cY,\cYb) \cD' \cY^\alpha \cDb \cYb^{\betab} + \cD' \cDb f(\cY,\cYb)~.
\end{align}
We will choose to leave this promotion of coordinates to superfields implicit rather than make the notation unreadable.

Returning now to the $\Psit$, we see that the space of fields satisfying all of the requirements is infinite dimensional.  To obtain a sensible description of the deformation space, we recall one more statement from conformal perturbation theory and Calabi-Yau NLSMs:  a supersymmetric deformation of the theory by a global D-term of the form
\begin{align}
\Delta S_D & = \int d^2 z \cD_{\text{tot}} f ~,
\end{align}
amounts to a shift of the K\"ahler potential by a global function.  We expect any such small perturbation to be marginally irrelevant, i.e. to lead to the same IR fixed point.  This fits well with the statement in conformal perturbation theory that a supersymmetric D-term deformation of a compact unitary (2,2) SCFT is necessarily irrelevant.\footnote{This is a well-known statement---see, for example,~\cite{Bertolini:2014ela,Gomis:2015yaa}.}

With this extra condition, we now observe  that if $\omega -\omega'= \p \pb f$ for any function $f \in \cA^{0,0} (\bY)$, then we expect $\omega$ and $\omega'$ to lead to the same IR fixed point.  Hence, the space of marginal (a,c) deformations is isomorphic to the quotient
\begin{align}
\{ \omega \in \cA^{1,1}_Y ~ | d\omega = 0\}/ \{ \omega = \p \pb f ~~| f \in \cA^{0,0}_Y\}~.
\end{align}
This is precisely the definition of the Bott-Chern cohomology group $H^{1,1}_{\text{BC}} (\bY, \C)$.\footnote{A useful review of various cohomology theories on a complex manifold is given in~\cite{Angella:2014ca}.}  Because $\bY$ is a compact Calabi-Yau space, it obeys the $\p\pb$ lemma, and that in turn implies the isomorphism
\begin{align}
H^{p,q}_{\text{BC}} (\bY, \C) \simeq H^{p,q}_{\pb} (\bY)~.
\end{align}
Taking the case of $p=q=1$ and using $H^{1,1}_{\pb} (\bY) \simeq H^{0,1}_{\pb} (\bY, \Omega^1_{\bY})$, we recover the expected result for $T_{\cM^{\text{ac}}}$.

In the same spirit, we now tackle the (c,c) deformation.  We seek fields $\Psi$ with the following properties:
\begin{enumerate}
\item $\Psi$ is well-defined on the NLSM targetspace and is expressed in terms of the superfields $\cY,\cYb$, and their superspace derivatives;
\item it has $\GUL\times\GUR$ charges $q_{\sleft} = 1$ and $q_{\sright} = +1$;
\item it carries (classical) dimensions $h_{\sleft} = h_{\sright} = \ff{1}{2}$;
\item $\Psi$ is chiral up to the NLSM equations of motion.
\end{enumerate}
The first three properties then require
\begin{align}
\Psi & =  \omega_{\alphab\betab} \cDb \cYb^{\alphab} \cDb' \cYb^{\betab} + \cDb \cDb' f~,
\end{align}
where $\omega \in \cA^{0,0}_Y(\bar{T}^\ast_{\bY} \otimes \bar{T}^\ast_{\bY})$, and $f \in \cA^{0,0}_Y$.  

Using the NLSM equations of motion, we find that the last requirement translates into two differential conditions that involve the K\"ahler connection, which we denote by $\nabla$:
\begin{align}
\nabla_{\alphab} \omega_{\gammab\betab} & = 
\nabla_{\gammab} \omega_{\alphab\betab} ~, &
\nabla_{\betab} \omega_{\alphab\gammab} & = \nabla_{\gammab} \omega_{\alphab\betab} ~.
\end{align}
To solve these conditions it is convenient to define $\eta^\beta_{\alphab} = \omega_{\alphab\betab} \cG^{\betab \alpha}$.  The first differential condition is then equivalent to $\pb \eta = 0$, while the second becomes
\begin{align}
\label{eq:mucondition}
\nabla_{\gammab} \mu_{\alphab\betab} & = 0~,
\end{align}
where $\mu \in \cA^{0,2}_Y$ is given by
\begin{align}
\mu_{\betab\alphab} & = \cG_{\beta\betab} \eta^{\beta}_{\alphab}-\cG_{\beta\alphab} \eta^{\beta}_{\betab}~.
\end{align}
The condition~(\ref{eq:mucondition}) is restrictive.  If we use the metric to raise the indices and contract with the unique holomorphic $d$-form $\Omega$, we obtain
\begin{align}
\mut_{\beta_1\cdots \beta_{d-2}} = \mu_{\alphab\betab} \cG^{\alphab \alpha} \cG^{\betab \beta} \Omega_{\alpha \beta \beta_1\cdots \beta_{d-2}}~,
\end{align}
and the condition on $\mu$ is equivalent to $\pb \mut = 0$, i.e. $\mut$ defines a class in $H^{d-2,0}_{\pb}(\bY,\cO_{\bY})$.  
This group is empty because $\bY$ is Calabi-Yau, which implies $\mut = 0$.  Because $\Omega$ is non-degenerate, we conclude that $\mu = 0$ as well.  So, the only way to satisfy our conditions is to solve
\begin{align}
\label{eq:etaconditions}
\pb \eta & = 0~, &
\cG_{\beta\betab} \eta^{\beta}_{\alphab}-\cG_{\beta\alphab} \eta^{\beta}_{\betab} & = 0~.
\end{align}
Let $\eta$ be a representative of a cohomology class $[\eta] \in H^{0,1}_{\pb} (\bY, T_{\bY})$.  We will now show that we can always find another representative
\begin{align}
\etat = \eta + \pb \lambda~
\end{align}
for some $\lambda \in \cA^{0,0}(\bY,T_{\bY})$ such that $\etat$ satisfies the second condition in~(\ref{eq:etaconditions}).

We need to find $\lambda$ such that
\begin{align}
\nabla_{\betab} \lambda^\beta \cG_{\beta \alphab} -
\nabla_{\alphab} \lambda^\beta \cG_{\beta\betab} & =\cG_{\beta\betab} \eta^{\beta}_{\alphab}-\cG_{\beta\alphab} \eta^{\beta}_{\betab}~.
\end{align}
Using $\pb \eta = 0$ it is not hard to show that the right-hand-side is a $\pb$-closed (0,2) form.  On $\bY$ any such form is $\pb$--exact, so that there exists some (0,1) form $\rho$ such that
\begin{align}
\cG_{\beta\betab} \eta^{\beta}_{\alphab}-\cG_{\beta\alphab} \eta^{\beta}_{\betab} =
\nabla_{\betab} \rho_{\alphab} - \nabla_{\alphab} \rho_{\betab}~.
\end{align}
We can therefore set $\lambda^\beta = \cG^{\betab\beta} \rho_{\betab}$.

We have shown that every cohomology class $[\eta] \in H^{0,1}_{\pb} (\bY, T_{\bY})$ has a representative $\eta$ satisfying~(\ref{eq:etaconditions}); we can change the representative to $\eta' = \eta + \pb \lambda$, which will also satisfy~(\ref{eq:etaconditions}) if and only if $\lambda$ obeys
\begin{align}
\nabla_{\betab} (\lambda^\beta \cG_{\beta\alphab}) - 
\nabla_{\alphab} (\lambda^\beta \cG_{\beta\betab})  = 0~.
\end{align}
On $\bY$ this is only possible if $\lambda^\beta = \nabla^\beta f$ for some function $f$.  Coming back to the form of the deformation, we see that such a shift amounts to $\omega_{\alphab\betab} \to \omega_{\alphab\betab} + \nabla_{\alphab}\nabla_{\betab} f$.  Using our equations of motion we have
\begin{align}
\cDb \cDb' f = \cDb \left[ \nabla_{\betab} f \cDb' \cYb^{\betab} \right]
= \p_{\alphab} \left(\nabla_{\betab} f\right) \cDb \cYb^{\alphab}\cDb' \cYb^{\betab}
+\nabla_{\betab}f\cDb \cDb'  \cYb^{\betab} = \nabla_{\alphab}\nabla_{\betab} f \cDb \cYb^{\alphab}\cDb' \cYb^{\betab}~.
\end{align}
So, the remaining freedom in shifting the representative of $[\eta]$ yields an irrelevant D-term deformation.  We have recovered the other familiar result:  $T_{\cM^{\text{cc}}} \simeq H^{0,1}_{\pb} (\bY, T_{\bY})$.

It is not surprising that we have reproduced the expected structure for the first order deformations of a large radius non-linear sigma model, because the supposition is that this Lagrangian theory is indeed superconformal for an appropriately chosen K\"ahler metric $\cG$.  We note that the two types of deformation differ in one important aspect:  we did not use the equations of motion in discussing the (a,c) deformations, and, indeed, there is no issue with adding to the action a small but finite (a,c) deformation of the form above.  This will shift the complexified K\"ahler class of the theory, and of course the action so obtained is equivalent to one with a new K\"ahler potential $\cK_{\text{new}}$.   On the other hand, the (c,c) deformation as we have written it is only infinitesimal because the supersymmetry requirements only hold up to equations of motion.  This is easy to understand:  the action written in terms of a choice of chiral superfields uses a fixed complex structure, so while the deformation is certainly integrable (either in the sense of complex geometry or superconformal field theory), we cannot hope to express the form of a finite deformation in terms of the original chiral superfields.

\subsection{A chiral algebra perspective}
Before we leave this warm-up exercise, we point out one more perspective that will be useful to us below, a view based on the chiral algebra of the theory, which we can think of as the cohomology of $\cDb$.\footnote{Foundational papers on this structure in two-dimensional theories include~\cite{Witten:1993jg,Silverstein:1995re,Witten:2005px}.  The structure has a close relationship to the chiral de Rham complex~\cite{Malikov:1998dw} and its generalizations.  It has been explored more recently in the context of Landau-Ginzburg models in~\cite{Dedushenko:2015opz} and in hybrid CFTs in~\cite{Bertolini:2013xga,Bertolini:2018now}.  A pedagogical discussion is given in~\cite{Melnikov:2019tpl}, and subtleties in (0,2) applications are pointed out in~\cite{Bertolini:2021hal}.}  This is a structure that exists in any (0,2) quantum field theory, and in favorable circumstances we can assume that it is isomorphic to the cohomology of the right-moving supercharge $G^+_{-1/2}$ of the IR SCFT.  In the case of (2,2) theories that we consider the $\cDb$ cohomology contains a holomorphic N=2 superconformal algebra that includes a representative of the $\GU(1)_{\sleft}$ current~\cite{Bertolini:2013xga}.  We will assume that in the IR this algebra indeed becomes the left-moving superconformal algebra. 

This offers a straightforward way to identify representatives of marginal (a,c) and (c,c) operators:  we need to merely identify the cohomology classes of operators with $q_{\sright} = +1$, $q_{\sleft} = \pm 1$, and spin $0$.  If our assumption about the RG flow is correct, then each such cohomology class corresponds in the SCFT to a chiral primary operator on the right with $q_{\sright} = +1$.  Since the RG flow preserves the spin $h_{\sleft}-h_{\sright} = 0$, we also have 
\begin{align}
h_{\sleft} = h_{\sright} = \ff{1}{2} q_{\sright} = \ff{1}{2}~.
\end{align}
Since $q_{\sleft} = \pm 1$, the operator must therefore either be anti-chiral primary on the left ($q_{\sleft} = -1$) or chiral primary on the left ($q_{\sleft} = +1$).  It is a simple exercise to apply this the (a,c) and (c,c) deformations of the classical non-linear sigma model to easily reproduce the results we reviewed above.  However, the point for us is that studying the chiral algebra will be much simpler in the massive theories that are our main interest.

Although we will not pursue this in this work, it is important to keep in mind that this identification is computationally powerful.  For example, it allows us to evaluate correlation functions and OPEs of these operators in a half-twisted theory, and these computations are essentially as powerful as similar computations in a topologically twisted theory~\cite{Adams:2005tc,Melnikov:2019tpl}, at least at genus $0$.

Denoting by $\cH_{\cDb}$ the full chiral algebra of the theory, we are then interested in characterizing the subspaces $\cH^{\text{ac}}_{\cDb}$ and $\cH^{\text{cc}}_{\cDb}$ corresponding to spin $0$ operators with $q_{\sleft}$, $q_{\sright}$ as described above.  These vector spaces have subspaces defined by the toric and polynomial deformations:
\begin{align}
\cH^{\text{toric}}_{\cDb} & \subseteq \cH^{\text{ac}}_{\cDb} ~,&
\cH^{\text{poly}}_{\cDb} & \subseteq \cH^{\text{cc}}_{\cDb}~.
\end{align}
Non-toric and non-polynomial deformations are then naturally thought of as equivalence classes belonging to, respectively, the quotient vector spaces $\cH^{\text{ac}}_{\cDb}/\cH^{\text{toric}}_{\cDb}$ and  $\cH^{\text{cc}}_{\cDb}/\cH^{\text{poly}}_{\cDb}$, and our goal is to provide an appropriate operator for each equivalence class.

\section{Hypersurface geometry} \label{s:Defsviageom}
In this section we use a geometric perspective to characterize the deformations for a special class of Calabi-Yau manifolds:  $X$ is a smooth hypersurface in a projective and simplicial $4$-dimensional NEF Fano toric variety $V$ with at worst terminal singularities.  Recall that a variety $V$ is NEF Fano if and only if it is complete, and its anti-canonical divisor $-K_V$ is NEF, i.e. has a non-negative intersection with every curve in $V$.  Moreover, $V$ is a Gorenstein variety, and its only singularities are terminal Gorenstein singularities which occur in co-dimension $4$~\cite{Cox:2000vi}: these singularities are missed by a generic hypersurface.  We focus on this class because it contains an enormous set of examples~\cite{Kreuzer:2000xy} with a simple combinatorial description, a canonical lift to a UV gauged linear sigma model~\cite{Morrison:1994fr}, and a beautiful mirror construction~\cite{Batyrev:1994hm}.   Moreover, there is a concrete description of the deformation spaces and their splits into toric/non-toric and polynomial/non-polynomial sets.  

The characterization of the deformations was crucial for early tests of mirror symmetry in this construction~\cite{Batyrev:1994hm,Aspinwall:1993rj}, but instead of giving the usual treatment---for example reviewed in~\cite{Cox:2000vi}---we will give a presentation that is well-suited for our purposes following a method given in~\cite{Aspinwall:2010ve}. 

\subsection{A little toric geometry}
We begin by setting notation and summarizing a few key results in toric geometry, mostly following the excellent text~\cite{Cox:2011tv}.

Fix a $d$-dimensional lattice $N \simeq \Z^d$.  Let $V$ be a projective simplicial toric variety with fan $\Sigma_V \subset N_{\R} = N\otimes_{\Z} \R$.  Denote by $\Sigma_V(1)$ the collection of $1$-dimensional cones, indexed by the primitive generators $u_\rho \in N$, with $\rho =1,\ldots, n = |\Sigma_V(1)|$.  In terms of the homogeneous Cox coordinates, for every $\rho$ there is a homogeneous coordinate $Z_\rho$ for $\C^n$, and we can describe $V$ as a quotient
\begin{align}
V = \left\{ \C^n \setminus F\right\} / \left\{ (\C^\ast)^{n-d}\times H\right\}~,
\end{align}
where $H$ is a finite abelian group, $F$ is a union of intersections of hyperplanes determined by the fan, and the $\C^\ast$ action is encoded in a matrix of charges $\bq$.  The toric divisors $D_\rho$, obtained as projections of the loci $\{\Z_\rho = 0\}$ will play an important role in our story.  We note two key properties:
\begin{enumerate}
\item the canonical divisor of the toric variety is given by
\begin{align}
K_V = -\textstyle\sum_\rho D_\rho~;
\end{align}
\item each toric divisor is Cartier, and the group of line bundles $\Pic(V)$ is generated by the corresponding line bundles $\cO_V(D_\rho)$, where $\cO_V$ is the structure sheaf of $V$.  We set $W = \Pic(V)\otimes_{\Z} \C$.
\end{enumerate}
The tangent sheaf $T_V$ and the cotangent sheaf $\Omega^1_V$ fit into the exact sequences\footnote{When $V$ is smooth, these sheaves have their usual geometric meaning.  More generally, when $V$ is a projective and simplicial, these sheaves should be understood as the appropriate generalizations of the geometric objects.   A careful discussion is given in~\cite{Cox:2011tv}.}
\begin{equation}
\label{eq:Euler}
\begin{tikzcd}
0 \ar[r] & W^\ast \otimes\cO_V \ar[r,"E"] &  \bigoplus_{\rho} \cO_V(D_\rho) \ar[r] &T_V \ar[r] & 0~, \\ 
0 \ar[r] & \Omega_V^1 \ar[r] &  \bigoplus_{\rho} \cO_V(-D_\rho) \ar[r,"E^T"] & W\otimes\cO_V \ar[r] & 0~, \\ 
\end{tikzcd}
\end{equation}
where the map $E$ is given by
\begin{align}
E(v) & = \left(v \cdot \bq_1 Z_1, v\cdot \bq_2 Z_2, \ldots, v\cdot \bq_n Z_n\right)~.
\end{align}
Using these exact sequences it is possible to prove a number of remarkable vanishing theorems that hold for NEF Fano simplicial toric varieties, including\footnote{Proofs and details of these theorems can be found in chapter 9 of~\cite{Cox:2011tv}.}
\begin{align}
\label{eq:CohomologyVanishing}
H^p(V,\Omega_V^q) = 0 \qquad \text{for $p \neq q$}~,
\end{align}
and for any NEF divisor $D$ on $V$
\begin{align}
\label{eq:DemazureVanishing}
H^p(V,\cO_V(D)) = 0 \qquad \text{for $p >0$}~.
\end{align}
We will often use these vanishing results together with Serre duality (which holds since $V$ is Gorenstein):
\begin{align}
H^p(V,\cE) \simeq \overline{H^{d-p} (V, \cE^\ast \otimes \cO_V(K_V))}~.
\end{align}
Using these results we can prove another vanishing result that will play an important role in what follows:
\begin{align}
\label{eq:MinusDrhoVanishing}
H^i(V, \cO_V(-D_\rho) ) = 0~.
\end{align}
This can be seen as follows.  First we observe that $H^0(V,\cO_V(-D_\rho)) = 0$ because given a section $\lambda \in H^0(V,\cO_V(-D_\rho))$ we would obtain a non-constant section $Z_\rho \lambda \in H^0(V,\cO_V)$, which is impossible on a projective variety.  Next, the cotangent sheaf exact sequence leads to a long exact sequence in cohomology which includes
\begin{equation}
\begin{tikzcd}
\cdots \ar[r] & H^{i} (V,\Omega_V^1) \ar[r] &  \bigoplus_{\rho} H^i(V,\cO_V(-D_\rho)) \ar[r] & H^i(V,W\otimes\cO_V) \ar[r] & \cdots, \\ 
\end{tikzcd}
\end{equation}
so that using~(\ref{eq:CohomologyVanishing}) we see $H^i (V, \cO_V(-D_\rho) ) =0$ for $i\ge 2$.  The remaining part of the long exact sequence is
\begin{equation}
\begin{tikzcd}
0 \ar[r] & H^{0} (V,W\otimes\cO_V) \ar[r] & H^1(V,\Omega_V^1) \ar[r] &  \bigoplus_{\rho} H^1(V,\cO_V(-D_\rho)) \ar[r] & 0~, \\ 
\end{tikzcd}
\end{equation}
but since the first two terms are isomorphic for a projective simplicial toric variety, the desired result holds for $i=1$ as well.

\subsection{Complex structure deformations}
We set $X = \{P = 0\} \subset V$, where $P$ is a generic holomorphic section of the anticanonical bundle:  $P \in H^0(V,\cO_V(-K_V))$.
We reviewed above that $T_{\cM^{\text{cc}}} \simeq H^1(X,T_X)$.  Our goal now is to describe $H^1(X,T_X)$ for the hypersurface in a way that explicitly identifies the polynomial and non-polynomial deformations.  In this section we closely follow~\cite{Aspinwall:2010ve}.  The first step is to observe that the adjunction sequence together with the Euler sequence of~(\ref{eq:Euler}) imply that the tangent sheaf $T_X$ is obtained as the cohomology of the complex
\begin{equation}
\label{eq:TangentComplex}
\begin{tikzcd}
\cE^{\bullet}  = 0 \ar[r] & W^\ast \otimes\cO_X \ar[r,"E"] & \underbrace{\bigoplus_\rho \cO_X(D_\rho)}_{= \cE^0} \ar[r,"dP"] & \cO_X(-K_B) \ar[r]  & 0~.
\end{tikzcd}
\end{equation}
This complex is exact except at the $0$-th position, and when $V$ is smooth it has the interpretation that vectors on $X$ are the vectors on $V$ that preserve the hypersurface.\footnote{This is familiar to gauged linear sigma model experts, making its appearance in that context already in~\cite{Witten:1993yc}.}  The sheaves on $X$ that show up in~(\ref{eq:TangentComplex}) are obtained by pulling back divisors from $V$ to $X$, and are related to sheaves on $V$ through the exact sequence
\begin{equation}
\label{eq:FromVtoX}
\begin{tikzcd}
0 \ar[r] &\cO_V(D + K_V) \ar[r] & \cO_V(D) \ar[r] &\cO_X(D) \ar[r] & 0~.
\end{tikzcd}
\end{equation}
The total cohomology, also known as hypercohomology, of the complex $\cE^{\bullet}$, calculated by a spectral sequence whose first page is $E_1^{p,q} = H^q(X,\cE^p)$, converges to $H^{p+q}(X,\cE^\bullet)$.  Since $T_X$ is obtained as the ordinary cohomology of $\cE^\bullet$, which fails to be exact just at the middle $\cE^0$ term, this gives a method for calculating $H^1(X,T_X)$.  In more detail, the first page of the spectral sequence only has non-zero entries for $ |p| \le 1$, which include
\begin{align*}
\begin{tikzpicture}
[zerogroup/.style={rectangle, fill = red!20, thin, inner sep = 0pt, minimum size = 4mm},scale = 1]
\node (H2OX) at (0,0) [zerogroup] {$H^2(X,W^\ast\otimes\cO_X)$};
\node (H2OXDr) at (5,0) [] {$\bigoplus_\rho H^2(X,\cO_X(D_\rho))$};
\node (H2OXKV) at (10,0) [zerogroup] {$ H^2(X,\cO_X(-K_V))$};
\draw[ thick,->] (H2OX) -- (H2OXDr);
\draw[thick,->] (H2OXDr) --(H2OXKV);
\node (H1OX) at (0,-1) [zerogroup] {$H^1(X,W^\ast\otimes\cO_X)$};
\node (H1OXDr) at (5,-1) [] {$\bigoplus_\rho H^1(X,\cO_X(D_\rho))$};
\node (H1OXKV) at (10,-1) [zerogroup] {$ H^1(X,\cO_X(-K_V))$};
\draw[ thick,->] (H1OX) -- (H1OXDr);
\draw[thick,->] (H1OXDr) --(H1OXKV);
\node (H0OX) at (0,-2) [] {$H^0(X,W^\ast\otimes\cO_X)$};
\node (H0OXDr) at (5,-2) [] {$\bigoplus_\rho H^0(X,\cO_X(D_\rho))$};
\node (H0OXKV) at (10,-2) [] {$ H^0(X,\cO_X(-K_V))$};
\draw[ thick,->] (H0OX) -- (H0OXDr);
\draw[thick,->] (H0OXDr) --(H0OXKV);
\draw[thin,->] (2.5,-3) -- (2.5,1) node[left] {$q$};
\draw[thin,->] (-2,-2.8) -- (12.5,-2.8) node[below] {$p$};
\draw[dashed,thin] (0,0) -- (10,-2);
\draw (0,-3) node {$p=-1$};
\draw (5,-3) node {$p=0$};
\draw (10,-3) node {$p=1$};
\end{tikzpicture}
\end{align*}
To obtain $H^1(X,T_X)$ we focus on the cohomology along the dashed line, and this is easily evaluated because the groups marked in light pink are zero.  The groups on the left are zero because $X$ is Calabi-Yau.  The groups on the right are zero because~(\ref{eq:FromVtoX}) implies $H^i(V,\cO_V(-K_V)) \simeq H^i(V,\cO_X(-K_V))$ for $i>0$, and the latter groups vanish by~(\ref{eq:DemazureVanishing})~.

The bottom row corresponds to deformations of the defining hypersurface equation, while the second row gives the non-polynomial deformations:
\begin{align}
T_{\cM^{\text{cc}}} / \left.T_{\cM^{\text{cc}}}\right|_{\text{poly}}  = \textstyle\bigoplus_\rho H^1(X,\cO_X(D_\rho))~.
\end{align}
While this result was given in~\cite{Aspinwall:2010ve}, we will next take it a step further and show
\begin{align}
T_{\cM^{\text{cc}}} / \left.T_{\cM^{\text{cc}}}\right|_{\text{poly}}  = \textstyle\bigoplus_\rho H^1(X,\cO_X(D_\rho)) = H^1(V,T_V)~,
\end{align}
i.e. the non-polynomial deformations of the hypersurface $X \subset V$ are exactly the deformations of complex structure of the ambient variety $V$---a satisfying result that was already obtained through a somewhat different approach in~\cite{Mavlyutov:2003fa}.

To prove the desired isomorphism we note that Serre duality and~(\ref{eq:MinusDrhoVanishing}) imply that $H^i(V,\cO_V(D_\rho+K_V)) = 0$ for all $i$. This in turn implies via~(\ref{eq:FromVtoX}) that $H^i(X,\cO_X(D_\rho)) = H^i(V,\cO_V(D_\rho))$~.  Finally, taking the long exact sequence associated to the Euler sequence for tangent sheaf on $V$, the result follows.

\subsection{Complexified K\"ahler deformations} \label{ss:CCdefs}
While the results in the previous section were a review of previous work, reproducing a  known result from a somewhat different point of view, we will now apply the same machinery to discussing the toric and non-toric complexified K\"ahler deformations, which to our best knowledge have not been previously considered from this point of view.

The conventional view on these deformations is obtained in three statements~\cite{Cox:2000vi}.  First, we observe that given the inclusion $i: X \hookrightarrow V$ we can pull back divisors on $V$ to those on $X$.  However, some divisors on $V$ do not intersect $X$, and these pull back to $0$ (up to linear equivalence) on $V$.  Taking this into account we obtain the toric divisors on $X$ and then of course also the corresponding classes in $H^{1}(X,\Omega_X^1)$.  Finally, it can be that some of the toric divisors become reducible when pulled back to $X$, leading to independent complexified K\"ahler deformations on $X$ that cannot be obtained by pulling back a complexified K\"ahler class from $V$.  

We will instead follow a different approach to describe the non-toric deformations directly in terms of properties of $V$.  The idea is simple:  we can apply exactly the methods of the previous section but now to the cotangent sheaf represented as the cohomology of the complex
\begin{equation}
\cF^{\bullet} = 
\begin{tikzcd}
0 \ar[r] & \cO_X(K_V) \ar[r,"dP"] &\underbrace{ \bigoplus_\rho \cO_X(-D_\rho)}_{=\cF^0} \ar[r,"E^T"] & W \otimes \cO_X \ar[r] & 0~. 
\end{tikzcd}
\end{equation}
The first page of the spectral sequence for the total cohomology of $\cF^{\bullet}$ is then
\begin{align*}
\begin{tikzpicture}
[zerogroup/.style={rectangle, fill = red!20, thin, inner sep = 0pt, minimum size = 4mm},scale = 1]
\node (H2KB) at (0,0) [zerogroup] {$H^2(X,\cO_X(K_V))$};
\node (H2OXDr) at (5,0) [] {$\bigoplus_\rho H^2(X,\cO_X(-D_\rho))$};
\node (H2OX) at (10,0) [zerogroup] {$ H^2(X,W\otimes \cO_X)$};
\draw[ thick,->] (H2KB) -- (H2OXDr);
\draw[thick,->] (H2OXDr) --(H2OXKV);
\node (H1KB) at (0,-1) [zerogroup] {$H^1(X,\cO_X(K_V))$};
\node (H1OXDr) at (5,-1) [] {$\bigoplus_\rho H^1(X,\cO_X(-D_\rho))$};
\node (H1OX) at (10,-1) [zerogroup] {$ H^1(X,W\otimes \cO_X)$};
\draw[ thick,->] (H1KB) -- (H1OXDr);
\draw[thick,->] (H1OXDr) --(H1OX);
\node (H0KB) at (0,-2) [zerogroup] {$H^0(X,\cO_X(K_V))$};
\node (H0OXDr) at (5,-2) [] {$\bigoplus_\rho H^0(X,\cO_X(-D_\rho))$};
\node (H0OX) at (10,-2) [] {$ H^0(X,W\otimes \cO_X)$};
\draw[ thick,->] (H1KB) -- (H1OXDr);
\draw[thick,->] (H1OXDr) --(H1OX);
\draw[thin,->] (2.5,-3) -- (2.5,1) node[left] {$q$};
\draw[thin,->] (-2,-2.8) -- (12.5,-2.8) node[below] {$p$};
\draw[dashed,thin] (0,0) -- (10,-2);
\draw (0,-3) node {$p=-1$};
\draw (5,-3) node {$p=0$};
\draw (10,-3) node {$p=1$};
\end{tikzpicture}
\end{align*}
Once again we marked the vanishing groups in light pink; this time it is the groups on the left that vanish by combining~(\ref{eq:FromVtoX}) with~ (\ref{eq:CohomologyVanishing}) and~(\ref{eq:DemazureVanishing}), while the groups on the right vanish because $X$ is Calabi-Yau. The bottom row encodes the toric deformations, while from the first row we obtain
\begin{align}
T_{\cM^{\text{ac}}}/\left.T_{\cM^{\text{ac}}}\right|_{\text{toric}}  = \textstyle\bigoplus_\rho H^1(X,\cO_X(-D_\rho)) = H^2(V,\Omega_V^1\otimes \cO_V(K_V))~.
\end{align}
The last isomorphism can be obtained in two steps.  First, the long exact sequence associated to~(\ref{eq:FromVtoX}) with $D = -D_\rho$ and the vanishing~(\ref{eq:MinusDrhoVanishing}) yield the isomorphism 
\begin{align}
H^1(X,\cO_X(-D_\rho)) = H^2(V,\cO_V(-D_\rho)\otimes\cO_V(K_V))~.
\end{align}
Next, taking the cotangent sheaf exact sequence and tensoring with $\cO_V(K_V)$ we obtain the exact sequence
\begin{equation}
\begin{tikzcd}
0 \ar[r] & \Omega_V^1 \otimes\cO_V(K_V)\ar[r] &  \bigoplus_{\rho} \cO_V(-D_\rho)\otimes\cO_V(K_V) \ar[r,"E^T"] & W\otimes\cO_V(K_V) \ar[r] & 0~, 
\end{tikzcd}
\end{equation}
and since by Serre duality $H^i(V,\cO_V(K_V)) = 0$ for $i \neq 4$, the associated long exact sequence yields the claimed isomorphism.

\section{Marginal operators in the hypersurface hybrid} \label{s:Hybriddefs}
In the previous section we obtained a characterization of the non-toric and non-polynomial deformations of a hypersurface $X \subset V$:
\begin{align}
T_{\cM^{\text{cc}}}/\left.T_{\cM^{\text{cc}}}\right|_{\text{poly}}  & = H^1(V,T_V)~,&
T_{\cM^{\text{ac}}}/\left.T_{\cM^{\text{ac}}}\right|_{\text{toric}}  &=H^2(V,\Omega_V^1\otimes \cO_V(K_V))~.
\end{align}
We will now use this characterization to find $\cDb$ cohomology classes in the hybrid theory that represent each type of deformation.  

We will make a stronger assumption that $V$ is smooth.  This is the simplest setting for hybrid theories, since then the base degrees of freedom can be described by a smooth non-linear sigma model.  The gauged linear sigma model suggests that it should be possible to extend the analysis to any simplicial NEF Fano toric variety $V$, but we will not pursue this extension here.  At any rate even assuming that $V$ is smooth leaves us with plenty of examples with non-toric and non-polynomial deformations.

\subsection{The Lagrangian of the hypersurface hybrid}
Let $L = \cO_V(K_V)$ and take $Y$ to be the total space of the line bundle $L$, with projection $\pi: Y \to V$.   The fibration gives us a way to construct the action patch by patch.  Suppose $\{\mathfrak{U}_a\}_{a\in I}$ is a cover for $V$, with $\mathfrak{U}_a \simeq \C^4$ with local holomorphic coordinates $u^i$ and their complex conjugates $\ub^{\ib}$.  We can then cover $Y$ with patches $\mathfrak{U}_a \times \C$, and denote the fiber coordinate by $\phi$.

The hybrid superfields are obtained by promoting the holomorphic coordinates just described to chiral superfields $U^i$ and $\Phi$, and their conjugates to anti-chiral superfields $\Ub^{\ib}$, $\Phib$.  To make a connection with the previous description we can set $\cY^0 = \Phi$ and $\cY^i = U^i$.  To specify the hybrid action~(\ref{eq:NLSMAction}) we choose a superpotential $\cW =\Phi P$, where $P$ is obtained by pulling back a section of the dual bundle $L^\ast$, and we observe that the geometry has a natural vector field $v = \phi \frac{\p}{\p\phi}$ which assigns charge $+1$ to $\Phi$ and leaves the $U^i$ invariant.  Thus, if we can pick a K\"ahler metric for which $v$ generates an isometry, the action will have a $\GU(1)_{\sleft}\times\GU(1)_{\sright}$ symmetry.  Moreover, the symmetry will be anomaly-free since by construction $Y$ has a trivial canonical bundle.

To describe the K\"ahler potential further, we pick a Hermitian metric on $L$, that is a positive section $h \in \cA_V^{0,0}(L^\ast \otimes \Lb^\ast)$.  The most general K\"ahler potential consistent with the isometry generated by $\cL_v$ is then
\begin{align}
\cK &= \cK(u,\ub, \cR)~,
\end{align}
where $\cR = \phi h(u,\ub) \phib$.  To leading order in the fiber direction 
\begin{align}
\cK = \phi h \phib + \cK_{\text{base}}(u,\ub) + O( \cR^2)~,
\end{align}
where $\cK_{\text{base}}$ is a K\"ahler potential for a K\"ahler metric on the base $V$.   

Using the metric $h$ we define the Chern connection $A = \p \log h$  on the bundle $L$, as well as its conjugate $\Ab = \pb \log h$.  These connections have Hermitian curvature $F \in \cA^{1,1}_V$, with
\begin{align}
\p A & = 0~,&
\pb \Ab & = 0~,&
\pb A & = - \p \Ab = F = F_{i\jb} du^i \wedge d\ub^{\jb}~,
\end{align}
where $F_{i\jb} = - \p_{\jb} A_{i}$ satisfies $\overline{F_{i\jb}} = F_{\ib j}$.  All of these pull back to $Y$, so that for example $\pi^\ast(h)$ gives a metric on the pullback bundle $L_{\text{v}} = \pi^\ast(L)$.  To keep the notation reasonably uncluttered we will not write the pullbacks explicitly in what follows unless it is likely to cause confusion.  


When derived from a linear sigma model the K\"ahler potential $\cK$ is determined in terms of a solution to $|\Sigma_V(1)| -\dim V$ algebraic equations on each affine patch, but we will not need the explicit details of this metric.  We remark that in keeping with the hybrid philosophy really any choice of smooth $\cK$ should do, but a canonical choice is not readily available for a general NEF Fano $V$.  If $V$ is Fano, then we can choose  $\cK = \phi h\phib + \cK_{\text{base}}(u,\ub)$ because it is possible to find a smooth metric $h$ so that the curvature $F$ has positive eigenvalues at every point on the base, and the resulting K\"ahler form is non-degenerate on $Y$.  However, for a general NEF (as opposed to ample) line bundle it is not possible to choose such a metric $h$~\cite{Demailly:1994}, and this simple K\"ahler potential will lead to a degenerate K\"ahler form.

Having set up this basic machinery, we will now construct representatives in the $\cDb$ cohomology  $\cH_{\cDb}$ of the hybrid theory for each of the (a,c) and (c,c) deformations identified above.  We will work at the level of classical field theory, but we expect our results to be robust at the level of the chiral algebra.

\subsection{The toric (a,c) deformations}
With a little more diagram chasing, it is not hard to see that the toric (a,c) deformations are described by the quotient $H^1(V,\Omega_V^1) / H^1(V,\Omega_V^1\otimes L)$, where the map $H^1(V,\Omega_V^1 \otimes L) \to H^1(V,\Omega_V^1)$ is simply multiplication by $P$.  To translate this to a statement in $\cDb$ cohomology we define the superfield
\begin{align}
\Theta & = \cDb' \cK_{\phi} = (\cK' + R\cK'') h \left(\cDb' \Phib + \left(\Ab_{\ib}  + \ff{1}{\cK' + \cR \cK''} \cK'_{\ib}\right) \cDb'\Ub^{\ib} \Phib\right)~.
\end{align}
This is useful because the equations of motion~(\ref{eq:EOMK}) imply
\begin{align}
\label{eq:DbarTheta}
\cDb \Theta & = -2m \cW_\phi = -2m P~.
\end{align}
Now the toric deformations are described as follows.  Given $[\omega] \in H^1(V,\Omega_V^1)$ with representative $\omega$, we set
\begin{align}
\cO^{\text{toric}} [\omega] = \omega_{i\jb} \cD' U^i \cDb\Ub^{\jb}~.
\end{align}
This is clearly $\cDb$-closed, and shifting $\omega$ by a $\pb$-exact form leads to a $\cDb$-exact shift of $\cO^{\text{ac}}_{\text{toric}} [\omega]$.  Thus, we have a well-defined map
\begin{align}
\cO^{\text{toric}} :  H^1(V,\Omega_V^1) \to \cH_{\cDb}~.
\end{align}
However, not all of these operators define non-trivial classes in $\cH_{\cDb}$:  given $[\lambda] \in H^1(V,\Omega_V^1\otimes L)$ with representative $\lambda$, we can construct a well-defined operator
\begin{align}
-\ff{1}{2m} \Theta \lambda_{i\jb} \cD' U^i \cDb\Ub^{\jb}~
\end{align}
which satisfies
\begin{align}
\cDb\left(- \ff{1}{2m} \Theta \lambda_{i\jb} \cD' U^i \cDb\Ub^{\jb}\right) = P \lambda_{i\jb} \cD' U^i \cDb\Ub^{\jb}~.
\end{align}
So, we characterize the toric deformations as a subset of the chiral algebra
\begin{align}
\cH_{\cDb}^{\text{toric}} = \left\{\cO^{\text{toric}}[\omega]~~ |~~ [\omega] \in  H^1(V,\Omega_V^1) / H^1(V,\Omega_V^1\otimes L)\right\}~.
\end{align}

\subsection{The non-toric (a,c) deformations}
We start with a class $[\xi] \in H^{1,2}_{\pb} (V,L)$ with representative $\xi$.   Since $P\xi \in \cA_V^{1,2}$ is a $\pb$-closed form, and since $H^{2}(V,\Omega_V^1) = 0$, there exists $\mu \in \cA_V^{1,1}$ such that 
\begin{align}
P\xi = \pb \mu~.
\end{align}
Any two solutions, say $\mu$ and $\mu'$, will differ by (a possibly trivial) toric deformation, i.e. $[\mu - \mu'] \in H^1(V,\Omega_V^1)$.

Given such a $\xi$, we would like to find a $\cDb$-closed local field $\cO[\xi]$ with the following properties:
\begin{enumerate}
\item it should be linear in $\xi$;
\item it should have spin $0$;
\item it should have $q_{\sleft} = -1$;
\item it should have $q_{\sright} = +1$;
\item it should transform trivially from patch to patch (i.e. it should be well-defined in field space).
\end{enumerate}
If we limit ourselves to fields constructed from the fundamental fields and their superspace derivatives, then these requirements have a unique solution of the form
\begin{align}
\cO_{\text{guess}} = ( \cDb'\Phib + \cdots) h \xi_{i\jb\kb} \cD' U^i \cDb \Ub^{\jb} \cDb \Ub^{\kb}~,
\end{align}
where $\cdots$ denotes connection terms that make the term in the parentheses transform covariantly with respect to bundle transformations.  Such an improvement is exactly provided by the $\Theta$ defined in the previous section, so that
\begin{align}
\cO_{\text{guess}} & = \Theta\xi_{i\jb\kb} \cD' U^i \cDb \Ub^{\jb} \cDb \Ub^{\kb}~.
\end{align}
Because $\xi$ is $\pb$-closed it follows that
\begin{align}
\cDb \cO_{\text{guess}}  = -2m P\xi_{i\jb\kb} \cD' U^i \cDb \Ub^{\jb} \cDb \Ub^{\kb} = \cDb \left( -4m \mu_{i\jb} \cD' U^i \cDb\Ub^{\jb}\right)~,
\end{align}
where the second equality follows from our observation $P\xi = \pb \mu$.  We conclude that the field $\cO_{\text{ac}}[\xi]$ defined by
\begin{align}
\cO_{\text{ac}}[\xi] = \Theta\xi_{i\jb\kb} \cD' U^i \cDb \Ub^{\jb} \cDb \Ub^{\kb} + 4m \mu_{i\jb} \cD' U^i \cDb\Ub^{\jb}
\end{align}
is $\cDb$-closed and is well-defined in $\cH^{\text{ac}}_{\cDb}/ \cH^{\text{toric}}_{\cDb}$.  In fact $\cO_{\text{ac}}[\xi]$ gives a well-defined map between the cohomology groups:
\begin{align}
\cO_{\text{ac}} : H^2(V, \Omega^1_V \otimes L) \to \cH^{\text{ac}}_{\cDb}/ \cH^{\text{toric}}_{\cDb}~.
\end{align}
It suffices to show that $\cO_{\text{ac}}[\pb \eta]$ is $\cDb$-exact.  When $\xi = \pb \eta$ we can set $\mu = P\eta$,
and, using again~(\ref{eq:DbarTheta}),
\begin{align}
\cO_{\text{ac}}[\pb \eta] & = \Theta \cDb \left( \eta_{i\jb} \cD' U^i \cDb\Ub^{\jb}\right) + 4m P \eta_{i\jb} \cD'U^i \cDb\Ub^{\jb} 
 = \cDb \left(-2\Theta \eta_{i\jb} \cD' U^i \cDb\Ub^{\jb} \right)~.
\end{align}
So, we have a well-defined injective map $H^2(V, \Omega^1_V \otimes L) \to \cH^{\text{ac}}_{\cDb}/\cH^{\text{toric}}_{\cDb}$, and we expect each of these $\cDb$-cohomology classes to correspond to an (a,c) non-toric deformation of the IR theory.

\subsection{The polynomial deformations}
These deformations are understood as deformations of the chiral superpotential, and the corresponding (c,c) field is simply
\begin{align}
\cO^{\text{poly}}[f] & = m\Phi f(\cU)~,
\end{align}
where $f \in H^0(V,L^\ast)$.  While the operator is obviously $\cDb$-closed and carries correct charges, some of these are also $\cDb$-exact, as we see from the zeroth row of the spectral sequence computation of the (c,c) deformations above, which characterizes the polynomial deformations as $H^0(V,L^\ast) / H^0(V,T_V)$.  The map $H^0(V,T_V) \to H^0(V,L^\ast)$ arises as follows.  Let $t\in \cA^{0,0}(T_V)$ be a holomorphic vector field.  Because $H^1(V,\cO_V)$ is empty the form $t \llcorner F = t^i F_{i\jb} d\ub^{\jb}$ is $\pb$-exact:  $t \llcorner F = \pb \eta$ for some $\eta\in \cA^{0,0}_V$, and there is a corresponding holomorphic vector field $\boldsymbol{t} \in \cA^{0,0}_Y (T_Y)$ given by
\begin{align}
\boldsymbol{t} & =  t^i\left( \ff{\p}{\p u^i} - A_i v\right)-\eta v ~,
\end{align}
where $v$ is the vertical holomorphic Killing vector $v =  \phi\ff{\p}{\p\phi}$.  It is now easy to see that the function 
$\boldsymbol{g} =  \boldsymbol{t} \llcorner \p \cW = \boldsymbol{t}^\alpha \cW_\alpha$ is holomorphic and of the form $\boldsymbol{g} = \phi g$ for a section $g\in H^0(V,L^\ast)$ given by
\begin{align}
g = -\eta P + t^i ( P_i - A_i P) ~.
\end{align}
The map $t \mapsto g$ is the desired map $H^0(V,T_V) \to H^0(V,L^\ast)$, and corresponding to this we have 
\begin{align}
\cO^{\text{poly}}[g] =m \Phi g & = \cDb \left( -\ff{1}{2}\boldsymbol{t}^\alpha \cDb' \cK_\alpha\right)~.
\end{align}


\subsection{The non-polynomial deformations}
The non-polynomial deformations can be understood in a more familiar geometric framework than the non-toric ones.  The total space of the line bundle $L \to V$ is a holomorphic manifold $Y$, and given a deformation of complex structure of the base $V$ we can ask the natural question whether the deformation can be lifted to a deformation of complex structure of $Y$.  Fortunately for us, the answer has been provided in a much wider setting in classic work from more than sixty years ago~\cite{Atiyah:1957xx}:  if $\tau$ represents a class in $H^1(V,T_V)$ and $F$ the curvature of the line bundle, we can construct $[\tau \llcorner F] \in H^2(V, \cO_V)$, and $\tau$ can be lifted to a deformation of complex structure of $Y$ if and only if $[\tau \llcorner F] = 0$.  Explicitly, if there exists $\xi \in \cA^{0,1}_V$ such that
\begin{align}
\label{eq:unobstructed}
\tau^i_{\kb} F_{i \jb} - \tau^i_{\jb} F_{i\kb} = \p_{\kb} \xi_{\jb} - \p_{\jb} \xi_{\kb}~,
\end{align}
then we define $\boldsymbol{\tau} \in \cA_Y^{0,1} ( T_Y)$ by\footnote{Note this is similar to our discussion of lifting a holomorphic vector $t$ to a holomorphic vector $\boldsymbol{t}$ on $Y$.}
\begin{align}
\boldsymbol{\tau}= \left( \xi_{\jb} v+ \tau^i_{\jb} \left( \ff{\p}{\p u^i} - A_i v\right) \right) \otimes d\ub^{\jb}~.
\end{align}
It is easy to check that $\pb \boldsymbol{\tau} = 0$.  In our case $\xi$ exists because $H^2(V,\cO_V) = 0$.  Moreover, if $\tau = \pb \rho$ for some $\rho \in \cA_V^{0,0}(T_V)$, then we can set $\xi_{\jb} = \rho^i F_{i\jb}$, and in this case $\boldsymbol{\tau}$ is $\pb$-exact:
\begin{align}
\label{eq:tauYexact}
\boldsymbol{\tau} &= \pb\boldsymbol{\rho}~,  & \boldsymbol{\rho}  &=  \rho^i \left(\ff{\p}{\p u^i} - A_i v \right)~.
\end{align}
Thus, we have a well-defined map on cohomology:  $H^1(V,T_V) \to H^1(Y,T_Y)$.

Using the map $\tau \mapsto \boldsymbol{\tau}$ and a little bit of foresight, we make an Ansatz for the corresponding (c,c) field:
\begin{align}
\cO_{\text{cc}}[\tau] & = \boldsymbol{\tau}^\alpha_{\betab}\, \cDb' \cK_\alpha \, \cDb \cYb^{\betab} + 2m\boldsymbol{f}~,
\end{align}
where $\boldsymbol{f}$ is a function on $Y$; we will choose $\boldsymbol{f}$ presently.  We then calculate, using~(\ref{eq:EOMK}),
\begin{align}
\cDb \cO_{\text{cc}}[\tau] & = 2m\left(\Phi f_{\betab} - \boldsymbol{\tau}^\alpha_{\betab} \cW_\alpha\right) \cDb\cYb^{\betab}~.
\end{align}
Since $\cW$ is a well-defined function on $Y$,
\begin{align}
\boldsymbol{\tau} \llcorner \p\cW & = \boldsymbol{\tau}^\alpha_{\betab} \cW_\alpha d\yb^{\betab} \in \cA_Y^{0,1}~,
\end{align}
and explicitly it is given by
\begin{align}
\boldsymbol{\tau} \llcorner \p\cW & = \phi  \lambda~,
\end{align}
where 
\begin{align}
\lambda & = \left(\xi_{\jb} P + \tau^i_{\jb} ( P_i - A_i P) \right) d\ub^{\jb} \in \cA^{0,1}_{V} (L^\ast)~
\end{align}
is $\pb$-closed by~(\ref{eq:unobstructed}).  Furthermore, ~(\ref{eq:DemazureVanishing}) implies $H^1(V, L^\ast) = 0$, which means $\lambda = \pb \sigma$ for some $\sigma \in \cA_V^{0,0} (L^\ast)$.  Putting these results together it follows that 
\begin{align}
\boldsymbol{\tau} \llcorner \p\cW &= \pb ( \phi \sigma)~,
\end{align} 
so that choosing $\boldsymbol{f} = \phi \sigma$ leads to a $\cDb$-closed field
\begin{align}
\cO_{\text{cc}}[\tau] & =  \boldsymbol{\tau}^\alpha_{\betab}\, \cDb' \cK_\alpha \, \cDb \cYb^{\betab} + 2m \boldsymbol{f}~.
\end{align}
The choice of $\sigma$ is ambiguous up to shifts by elements of $H^0(V,T_V^\ast)$.  Just as in the preceding discussion of the non-toric deformations, this means $\cO_{\text{cc}}[\tau]$ is well-defined in the quotient $\cH^{\text{cc}}_{\cDb}/\cH^{\text{poly}}_{\cDb}$.

It remains to show that this gives a well-defined map on cohomology:
\begin{align}
\cO_{\text{cc}} :  H^1(V,T_V) \to \cH^{\text{cc}}_{\cDb}/\cH^{\text{poly}}_{\cDb}~,
\end{align}
and it suffices to check that $\cO_{\text{cc}}[\pb\rho]$ is $\cDb$-exact.   But, since $\tau =\pb\rho$ implies $\boldsymbol{\tau} = \pb \boldsymbol{\rho}$, we set $\boldsymbol{f} = \boldsymbol{\rho} \llcorner \p \cW = \boldsymbol{\rho}^\alpha \cW_\alpha$, and using again~(\ref{eq:EOMK})
\begin{align}
\cDb \left( -\boldsymbol{\rho}^\alpha \cDb' \cK_\alpha\right) = \boldsymbol{\tau}^\alpha_{\betab}\, \cDb' \cK_\alpha \, \cDb \cYb^{\betab}-\boldsymbol{\rho}^\alpha \cDb\,\cDb'\cK_\alpha =\cO_{\text{cc}} [\pb\tau]~.
\end{align}

\section{The NS--R sector of the hypersurface hybrid} \label{s:HybridNSR}

We now discuss the computation of the marginal deformations in the NS--R sector of the hybrid theory.  That is, assuming the hybrid theory flows to a compact SCFT, we know that every right-moving chiral primary operator has an image as a right-moving ground state in the NS-R sector.  Using the technology developed in~\cite{Bertolini:2013xga} it is possible to compute all the states that correspond to massless spacetime fermions in a string compactification based on the SCFT.  A subset of these states, those with $q_{\sleft} = \pm 1$ and $h_{\sleft} = 1/2$, is isomorphic to the marginal deformations in the NS--NS sector.  Since the left-moving weights can be calculated using the chiral algebra, this gives an effective way to check our results and to also check that the techniques of~\cite{Bertolini:2013xga} really do apply to hypersurface hybrids.

We will see that the deformations are captured by the cohomology of the right moving supercharge $\overline{\mathbf{Q}}$, which in the hybrid decomposes into the sum of two anticommuting operators: $\overline{\mathbf{Q}}_0$, the supercharge of the base NLSM, and $\overline{\mathbf{Q}}_W$, the supercharge contribution from the inclusion of the superpotential $\mathcal{W} = \Phi P$. The (2,2) superfields are decomposed into their (0,2) components,

\begin{align}
    \mathcal{Y}^{\alpha} &= Y^{\alpha} + \sqrt{2} \theta' \mathcal{X}^{\alpha} + \theta' \overline{\theta}' \partial Y^{\alpha}, & \qquad
    \overline{\mathcal{Y}}^{\overline{\alpha}} &= \overline{Y}^{\overline{\alpha}} - \sqrt{2} \overline{\theta}' \overline{\mathcal{X}}^{\overline{\alpha}} - \theta' \overline{\theta}' \partial \overline{Y}^{\overline{\alpha}} \nonumber \\
    Y^{\alpha} &= y^{\alpha} + \sqrt{2} \theta \eta^{\alpha} + \theta \overline{\theta} \bar{\partial} y^{\alpha}, & \qquad
    \overline{Y}^{\alpha} &= \overline{y}^{\overline{\alpha}} - \sqrt{2} \overline{\theta} \overline{\eta}^{\overline{\alpha}} - \theta \overline{\theta} \bar{\partial} \overline{y}^{\overline{\alpha}} \nonumber \\
    \mathcal{X}^{\alpha} &= \chi^{\alpha} + \sqrt{2} \theta H^{\alpha} + \theta \overline{\theta} \bar{\partial} \chi^{\alpha}, & \qquad
    \overline{\mathcal{X}}^{\overline{\alpha}} &= \overline{\chi}^{\overline{\alpha}} + \sqrt{2} \overline{\theta} \overline{H}^{\alpha} - \theta \overline{\theta} \bar{\partial} \overline{\chi}^{\overline{\alpha}}
\end{align}

We identify $y^{\alpha}$ as coordinates on the total space $Y$ and decompose these into $(\phi,u^i)$ as above.  For the other component fields, we denote the fiber component with a $0$ superscript or subscript. We use the equations of motion  to eliminate the auxiliary fields and then make the following field redefintions,
\begin{equation}
    \overline{\chi}_{\alpha} = \cG_{\alpha \overline{\beta}} \overline{\chi}^{\overline{\beta}}, \qquad \rho_{\alpha} = \cG_{\alpha \overline{\alpha}} \partial y^{\alpha} + \Gamma^{\delta}_{\alpha \beta} \overline{\chi}_{\delta} \chi^{\beta}~.
\end{equation}
As described in~\cite{Bertolini:2013xga}, in the large radius limit these degrees of freedom can be treated as a free curved $bc$--$\beta\gamma$ system, while the right-moving degrees of freedom are taken in their ground states.


A general state in the $\overline{\textbf{Q}}$ cohomology in the NS--R sector is represented of a $(0,k)$ form $\Psi$ valued in a product bundle $(T^*_Y)^{\otimes s} \otimes (T_Y)^{\otimes t}$ contracted into $k$ copies of the zero modes $\overline{\eta}^{\bar{i}}$ and a combination of $\chi^{\alpha}$, $\overline{\chi}_{\alpha}$, and $\rho_{\alpha}$.  
The charges and weights of the fields are given in the table below, and this allows us to select the states that correspond to the marginal deformations.

\begin{table}[h]
\centering
\begin{tabular}{|l|l|l|l|l|l|l|l|l|}
\hline
 & $u^i$ & $\rho_i$ & $\chi^{i}$ & $\overline{\chi}_{i}$ & $\phi$ & $\rho_0$ & $\chi^0$ & $\overline{\chi}_0$ \\ \hline
$q_{\sleft}$            & $0$ & $0$ & $-1$ & $1$ & $1$ & $-1$ & $0$ & $0$  \\ \hline
$q_{\sright}$ & $0$ & $0$ & $0$  & $0$ & $1$ & $-1$ & $1$ & $-1$ \\ \hline
$2h_{\sleft}$                    & $0$ & $2$ & $1$  & $1$ & $1$ & $1$  & $2$ & $0$  \\ \hline
\end{tabular}
\end{table}

\subsection{(a,c) Deformations}
Given these, we construct the most general set of operators obeying our charge and weight constraints.  The corresponding states are then constructed by letting these act on the NS--R Fock vacuum.  Starting with the (a,c) deformations, we find operators and the eigenvalue of their coefficients $\Psi$ under the Lie derivative $\mathcal{L}_v$ that can contribute to NS--R states with $q_{\sleft} = -1$:
\begin{align}
\mathcal{O}^2 &= \Psi^2_{\alpha} \chi^{\alpha}~,&
\mathcal{O}^5 &= \Psi^{5,\beta}_{\alpha} \overline{\chi}_{\beta} \chi^{\alpha}~,&
\mathcal{O}^6 &= \Psi^{6,\beta} \rho_{\beta} + \Psi^{6,\beta}_{\alpha} \overline{\chi}_{\beta} \chi^{\alpha}~,&
\mathcal{O}^7 &= \Psi^{7,\alpha \beta} \rho_{\alpha} \overline{\chi}_{\beta}~, \nonumber\\
\cL_v \Psi^2 & = 0~,&
\cL_v \Psi^5 & =- \Psi^5~,&
\cL_v \Psi^6 & = - \Psi^6~,&
\cL_v \Psi^7 & = -2 \Psi^7~.
\end{align}
The coefficients $\Psi$ are sections of the following bundles,
\begin{align}
    \Psi^2 &\in \cA_Y^{0,u}(T^*_Y)~,& 
    \Psi^5 &\in \cA_Y^{0,u}(T^*_Y \otimes T_Y)~,&
    \Psi^6 &\in \cA_Y^{0,u}(T_Y)~,& 
     \Psi^7 &\in \cA_Y^{0,u}(T_Y \otimes T_Y)~.
\end{align}

By utilizing the decomposition $\overline{\mathbf{Q}} = \overline{\mathbf{Q}}_0 + \overline{\mathbf{Q}}_W$, we can compute the cohomology via a spectral sequence with zeroth stage $d_0 = \overline{\mathbf{Q}}_0$ and first stage $d_1 = \overline{\mathbf{Q}}_W$.
We start by taking the $d_0 = \overline{\mathbf{Q}}_0$ cohomology, which acts on the component fields as $\overline{\mathbf{Q}}_0 = -\overline{\eta}^{\overline{i}} \frac{\partial}{\partial \overline{u}^{\overline{i}}}$ and thus forces the $\Psi$ into cohomology groups $H^u(Y,(T^*_Y)^{\otimes s} \otimes (T_Y)^{\otimes t})$, which we will shorten on the diagram to $H^u(B_{s,t})$. To organize these states into a complex, we define $p = q_{\sright} - u$, and parameterize the complex by $(p,u)$. Below we have the first page of the spectral sequence,
\begin{equation*}
\begin{tikzpicture}
\draw[help lines, color=gray!30, dashed] (-4.9,-3);
\draw[->,thin] (-5,-2)--(8.3,-2) node[right]{$p$};
\draw[->,thin] (8,-2.3)--(8,3.3) node[above]{$u$};
\draw [dashed,thin] (-3.5,3) -- (6,0);
\put(-120,70){$0$}  \put(-95,70){$H^3(Y,B_{0,2})$} \put(-25,70){$H^3(Y,B_{1,1}) \oplus H^3(Y,B_{0,1})$} \put(120,70){$H^3(Y,B_{1,0})$} \put(200,70){$0$}
\put(-120,35){$0$}  \put(-95,35){$H^2(Y,B_{0,2})$} \put(-25,35){$H^2(Y,B_{1,1}) \oplus H^2(Y,B_{0,1})$} \put(120,35){$H^2(Y,B_{1,0})$} \put(200,35){$0$}
\put(-120,0){$0$}  \put(-95,0){$H^1(Y,B_{0,2})$} \put(-25,0){$H^1(Y,B_{1,1}) \oplus H^1(Y,B_{0,1})$} \put(120,0){$H^1(Y,B_{1,0})$} \put(200,0){$0$}
\put(-120,-35){$0$} \put(-95,-35){$H^0(Y,B_{0,2})$} \put(-25,-35){$H^0(Y,B_{1,1}) \oplus H^0(Y,B_{0,1})$} \put(120,-35){$H^0(Y,B_{1,0})$} \put(200,-35){$0$}
\end{tikzpicture}
\end{equation*}
The states we are after correspond to $q_{\sright} = 1$, which is denoted by the dashed line.
Before applying the $d_1$ stage of the sequence, we make use of the sheaf cohomology results developed in appendix C of \cite{Bertolini:2013xga}. Given a section of a bundle $\mathcal{E}$ on $Y$ at fixed grade $r$ in $\phi$, we are able to find an isomorphic bundle on $V$ in cohomology.
Each of the relevant bundles on $Y$ will have a corresponding exact sequence relating the base and fiber components of the bundle. For instance, consider the sequence for $T^*_Y$,
\begin{equation}
\begin{tikzcd}
    0 \arrow[r] & (\pi^*(T^*_V))_r  \arrow[r] & (T_Y^*)_r \arrow[r] & (\pi^*(L^{-1}))_{r-1}\arrow[r] & 0
\end{tikzcd}
\end{equation}
We wish to evaluate this sequence at grade $r = 0$, i.e. the eigenvalue of $\Psi^2$ under $\mathcal{L}_v$. This tells us that $T^*_Y$ at grade $0$ is equivalent to the pullback of $T^*_V$, which after taking cohomology allows us to use the isomorphism
\begin{equation}
    H^{\bullet}_r(Y, \pi^*(\mathcal{E})) \simeq H^{\bullet}(B, \mathcal{E} \otimes L^{-r})
\end{equation}
to obtain
\begin{equation}
    H^{\bullet}_{0}(Y,T^*_Y) = H^{\bullet}(V,T^*_V)
\end{equation}
We carry this exercise out for the remaining bundles, and are able to reduce the first page of the spectral sequence to
\begin{equation*}
\begin{tikzpicture}
\draw[help lines, color=gray!30, dashed] (-4.9,-3);
\draw[->,thin] (-5,-2)--(7.3,-2) node[right]{$p$};
\draw[->,thin] (7,-2.3)--(7,3) node[above]{$u$};
\put(-120,70){$0$}  \put(-95,70){$H^3(V,L^2)$} \put(-30,70){$H^3(V,T^*_V \otimes L) \oplus H^3(V,L)$} \put(115,70){$H^3(V,T^*_V)$} \put(180,70){$0$}
\put(-120,35){$0$}  \put(-95,35){$H^2(V,L^2)$} \put(-30,35){$H^2(V,T^*_V \otimes L) \oplus H^2(V,L)$} \put(115,35){$H^2(V,T^*_V)$} \put(180,35){$0$}
\put(-120,0){$0$}  \put(-95,0){$H^1(V,L^2)$} \put(-30,0){$H^1(V,T^*_V \otimes L) \oplus H^1(V,L)$} \put(115,0){$H^1(V,T^*_V)$} \put(180,0){$0$}
\put(-120,-35){$0$} \put(-95,-35){$H^0(V,L^2)$} \put(-30,-35){$H^0(V,T^*_V \otimes L) \oplus H^0(V,L)$} \put(115,-35){$H^0(V,T^*_V)$} \put(180,-35){$0$}
\end{tikzpicture}
\end{equation*}
From here, we can make use of various vanishing theorems and Serre duality for the cohomology of vector bundles on toric varieties. The new complex then takes the form
\begin{equation*}
\begin{tikzpicture}
\draw[help lines, color=gray!30, dashed] (-4,-3);
\draw[->,thin] (-5,-2)--(6.3,-2) node[right]{$p$};
\draw[->,thin] (6,-2.3)--(6,3) node[above]{$u$};
\put(-120,70){$0$}  \put(-80,70){$0$} \put(-35,70){$H^3(V,T^*_V \otimes L)$} \put(90,70){$0$} \put(130,70){$0$}
\put(-120,35){$0$}  \put(-80,35){$0$} \put(-35,35){$H^2(V,T^*_V \otimes L)$} \put(90,35){$0$} \put(130,35){$0$}
\put(-120,0){$0$}  \put(-80,0){$0$} \put(-35,0){$H^1(V,T^*_V \otimes L)$} \put(75,0){$H^1(V,T^*_V)$} \put(130,0){$0$}
\put(-120,-35){$0$} \put(-80,-35){$0$} \put(-55,-35){$H^0(V,T^*_V \otimes L) \oplus H^0(V,L)$} \put(90,-35){$0$} \put(130,-35){$0$}
\end{tikzpicture}
\end{equation*}
Ultimately, we are interested in the cohomology along the diagonal corresponding to $q_{\sright}=1$. So, we can zoom into the relevant parts and consider the $d_1 = \overline{\mathbf{Q}}_W$ map,
\begin{equation*}
\begin{tikzpicture}
\draw[help lines, color=gray!30, dashed] (-4.9,1);
\draw[->,thin] (-2,-1.5)--(6,-1.5);
\draw[->,thin] (5.5,-2)--(5.5,1);
\put(-50,10){$H^2(V,T^*_V \otimes L)$} \put(115,10){$0$}
\put(-50,-30){$H^1(V,T^*_V \otimes L)$} \put(100,-30){$H^1(V,T^*_V)$}
\draw[->] (2,.5)--(3,.5)node[above,xshift=-.5cm]{$\overline{\textbf{Q}}_W$};
\draw[->] (2,-.9)--(3,-.9)node[above,xshift=-.5cm]{$\overline{\textbf{Q}}_W$};;
\end{tikzpicture}
\end{equation*}
%
The $\overline{\textbf{Q}}_W$ action on the component fields is given by
\begin{equation}
    \overline{\textbf{Q}}_W \cdot \overline{\chi}_{\alpha} = W_{\alpha}, \quad \overline{\textbf{Q}}_W \cdot \rho_{\alpha} = \chi^{\beta} W_{\beta \alpha}
\end{equation}
and thus the $\overline{\textbf{Q}}_W$ map quotients out elements of $H^1(V,T^*_V \otimes L)$ multiplied by $P$ from $H^1(V,T^*_V)$. and acts by $0$ on $H^2(V,T^*_V \otimes L)$. So, the terminal stage of the spectral sequence converges to the cohomology of $\overline{\mathbf{Q}}$, giving
\begin{equation*}
\begin{tikzpicture}
\draw[help lines, color=gray!30, dashed] (-4.9,-3);
\draw[->,thin] (-2,-2)--(5.5,-2);
\draw[->,thin] (5,-2.5)--(5,1);
\put(-40,0){$H^2(V,T^*_V \otimes L)$} \put(100,0){$0$}
\put(-40,-40){$H^1(V,T^*_V \otimes L)$} \put(75,-40){$\frac{H^1(V,T^*_V)}{\Psi^2 \sim \Psi^2 + \Psi^{5} P}$}
\end{tikzpicture}
\end{equation*}
The bottom right corner is isomorphic to the toric deformations, while the top left gives the non-toric deformations as $H^2(V,T^*_V \otimes L)$, matching our result in section \ref{ss:CCdefs}.

\subsection{(c,c) Deformations}
This analysis is easily extended to the (c,c) deformations. We now search for all operators with $\GU(1)_{\sleft}\times\GU(1)_{\sright}$ charges $(1,1)$ and weight $h=\frac{1}{2}$. The full list and their $\phi$ grading are given below.
\begin{align}
    \mathcal{O}^{1,0} &= \Psi^{1,0}, \quad & \cL_v {\Psi^{1,0}} &= \Psi^{1,0}, \quad & \Psi^{1,0} &\in \cA^{0,u}_Y \nonumber \\
    \mathcal{O}^{1,1} &= \Psi^{1,1 \alpha} \overline{\chi}_{\alpha}, & \cL_v {\Psi^{1,1}} &= 0, & \Psi^{1,1} &\in \cA^{0,u}(T_Y) \\
    \mathcal{O}^{1,2} &= \Psi^{1,2 \alpha \beta} \overline{\chi}_{\alpha} \overline{\chi}_{\beta}, & \cL_v {\Psi^{1,2}} &= -\Psi^{1,2}, & \Psi^{1,2} &\in\cA_Y^{0,u}( \wedge^2 T_Y )\nonumber
\end{align}
The first stage of the spectral sequence gives the complex

\begin{tikzpicture}
\draw[help lines, color=gray!30, dashed] (-4.9,-3);
\draw[->,thin] (-4,-2)--(7.5,-2) node[right]{$p$};
\draw[->,thin] (7,-2.5)--(7,4) node[above]{$u$};
\draw [dashed,thin] (-4,2.6) -- (5.5,-1.5);
\put(-100,70){$0$}  \put(-65,70){$H^3(Y,\wedge^2 T_Y)$} \put(25,70){$H^3(Y,T_Y)$} \put(100,70){$H^3(Y,\mathcal{O}_Y)$} \put(170,70){$0$}
\put(-100,35){$0$}  \put(-65,35){$H^2(Y,\wedge^2 T_Y)$} \put(25,35){$H^2(Y,T_Y)$} \put(100,35){$H^2(Y,\mathcal{O}_Y)$} \put(170,35){$0$}
\put(-100,0){$0$}  \put(-65,0){$H^1(Y,\wedge^2 T_Y)$} \put(25,0){$H^1(Y,T_Y)$} \put(100,0){$H^1(Y,\mathcal{O}_Y)$} \put(170,0){$0$}
\put(-100,-35){$0$} \put(-65,-35){$H^0(Y,\wedge^2 T_Y)$} \put(25,-35){$H^0(Y,T_Y)$} \put(100,-35){$H^0(Y,\mathcal{O}_Y)$} \put(170,-35){$0$}
\end{tikzpicture}

\noindent where again we are interested in the cohomology along the dashed line at $q_{\sright}=1$. Applying the same isomorphisms to this complex and utilizing Serre duality gives
\begin{equation*}
\begin{tikzpicture}
\draw[help lines, color=gray!30, dashed] (-5,0);
\draw[->,thin] (-2,-2)--(8,-2);
\put(-20,0){$H^1(V,T_V)$} \put(145,0){$0$}
\put(-50,-40){$H^0(V,T_V) \oplus H^0(V,\mathcal{O}_V)$} \put(125,-40){$H^0(V,L^*)$}
\draw[->] (3,0)--(4,0)node[above,xshift=-.5cm]{$\overline{\textbf{Q}}_W$};
\draw[->] (3,-1.3)--(4,-1.3)node[above,xshift=-.5cm]{$\overline{\textbf{Q}}_W$};;
\end{tikzpicture}
\end{equation*}
The $\overline{\mathbf{Q}}_W$ map acts as before, and on the bottom row quotients out elements of $H^0(V,L^*)$ proportional to $\p\cW$. The $d_2$ map would take every entry into an empty group, so the spectral sequence already converges at this stage, and the cohomology is given by
\begin{equation*}
\begin{tikzpicture}
\draw[help lines, color=gray!30, dashed] (-4.9,1);
\draw[help lines, color=gray!30, dashed] (-4.9,-2.5);
\draw[->,thin] (-1,-2)--(6,-2);
\put(-10,-10){$H^1(V,T_V)$} \put(80,-10){$0$}
\put(-10,-40){$H^0(V,T_V)$} \put(60,-40){$\frac{H^0(V,L^*)}{\Psi^{1,0} \sim \Psi^{1,0} + dW \lrcorner \Psi^{1,1}}$}
\end{tikzpicture}
\end{equation*}
The bottom right gives the polynomial deformations, while the non-polynomial deformations are given in the top left by $H^1(V,T_V)$, as expected.

\section{Further directions} \label{s:Further}

In this paper we investigated a UV Lagrangian theory --- the hypersurface hybrid, which is expected to a compact (2,2) SCFT.  We established two main results.  First, we obtained explicit representatives for all marginal operators of the SCFT in terms of cohomology classes in the chiral algebra $\cH_{\cDb}$.  Second, we demonstrated that although the hypersurface hybrid is a rather degenerate example of a hybrid theory, nevertheless the hybrid methodology can be used to study its NS--R sector.

While our results certainly settle some questions of principle, they also bear on practical matters.  First, our construction of representatives of the non-toric and non-polynomial marginal operators could be used to evaluate correlation functions of all marginal operators in the theory, and perhaps they could already play a role in the mathematical formulations of topological field theories as in~\cite{Li:2019,Guffin:2008kt,Chen:2019wub}.  This will require an analysis of quantum corrections to our results, which are probably best analyzed in the language of the curved $bc$--$\beta\gamma$ system that encodes the chiral algebra of the hybrid theory~\cite{Bertolini:2013xga}.   It seems reasonable to conjecture that the quantum corrections to the form of the operators we gave can be absorbed into a redefinition of the K\"ahler potential $\cK$, which made its appearance in the construction through the superfield $\Theta$.  Of course there would be non-perturbative corrections to OPE of the (a,c) operators, and it would be extremely interesting to understand these directly in terms of the hybrid theory.  The results obtained recently in~\cite{Erkinger:2022sqs} should be of use in uncovering these quantum corrections.

More importantly, the work is a step towards a more ambitious UV lift of the infinitesimal deformations to the gauged linear sigma model.  Our construction of the operators was given in a particular large radius phase of the GLSM, and it relied on a number of geometric properties of this phase.  It would be interesting to study the chiral algebra of the GLSM in detail in order to find the non-toric and non-polynomial deformations in that UV description.  We hope that our explicit representatives might serve as a guide to finding that structure, which by its nature will be more combinatoric rather than geometric and will require some further developments of the gauged linear sigma model's chiral algebra.  Whatever the motivation for those explorations, a systematic understanding of the latter will be of great use to future generations of linear sigma model experts.

\bibliographystyle{utphys}
\bibliography{./newref}

\providecommand{\href}[2]{#2}\begingroup\raggedright\begin{thebibliography}{10}

\bibitem{Nemeschansky:1986yx}
D.~Nemeschansky and A.~Sen, ``{Conformal invariance of supersymmetric sigma
  models on Calabi-Yau manifolds},''
\href{http://dx.doi.org/10.1016/0370-2693(86)91394-8}{{\em Phys.Lett.}
  {\bfseries B178} (1986) 365}.

\bibitem{Polchinski:1998rr}
J.~Polchinski, {\em String Theory}, vol.~2.
\newblock Cambridge University Press, Cambridge, UK, 1998.

\bibitem{Greene:1997my}
B.~R. Greene, ``{String Theory on Calabi-Yau Manifolds},''
  \href{http://arxiv.org/abs/hep-th/9702155}{{\ttfamily arXiv:hep-th/9702155}}.

\bibitem{Hori:2003ds}
K.~Hori, S.~Katz, A.~Klemm, R.~Pandharipande, R.~Thomas, C.~Vafa, R.~Vakil, and
  E.~Zaslow, {\em Mirror symmetry}, vol.~1 of {\em Clay Mathematics
  Monographs}.
\newblock American Mathematical Society, Providence, RI, 2003.
\newblock With a preface by Vafa.

\bibitem{Witten:1993yc}
E.~Witten, ``{Phases of N = 2 theories in two dimensions},'' {\em Nucl. Phys.}
  {\bfseries B403} (1993) 159--222,
\href{http://arxiv.org/abs/hep-th/9301042}{{\ttfamily arXiv:hep-th/9301042}}.

\bibitem{Melnikov:2019tpl}
I.~V. Melnikov, ``{An Introduction to Two-Dimensional Quantum Field Theory with
  (0,2) Supersymmetry},''
\href{http://dx.doi.org/10.1007/978-3-030-05085-6}{{\em Lect. Notes Phys.}
  {\bfseries 951} (2019) }.

\bibitem{Candelas:1987kf}
P.~Candelas, A.~M. Dale, C.~A. Lutken, and R.~Schimmrigk, ``{Complete
  Intersection Calabi-Yau Manifolds},''
  \href{http://dx.doi.org/10.1016/0550-3213(88)90352-5}{{\em Nucl. Phys. B}
  {\bfseries 298} (1988) 493}.

\bibitem{Green:1987rw}
P.~Green and T.~Hubsch, ``{Polynomial deformations and cohomology of Calabi-Yau
  manifolds},''
{\em Commun. Math. Phys.} {\bfseries 113} (1987) 505.

\bibitem{Hubsch:1992nu}
T.~Hubsch, {\em {Calabi-Yau manifolds: A Bestiary for physicists}}.
\newblock World Scientific, Singapore, 1992.

\bibitem{Batyrev:1994hm}
V.~V. Batyrev, ``{Dual polyhedra and mirror symmetry for Calabi-Yau
  hypersurfaces in toric varieties},'' {\em J. Alg. Geom.} {\bfseries 3} (1994)
  493--545,
\href{http://arxiv.org/abs/arXiv:alg-geom/9310003}{{\ttfamily
  arXiv:alg-geom/9310003}}.

\bibitem{Batyrev:1994pg}
V.~V. Batyrev and L.~A. Borisov, ``{On Calabi-Yau complete intersections in
  toric varieties},''
\href{http://arxiv.org/abs/alg-geom/9412017}{{\ttfamily
  arXiv:alg-geom/9412017}}.

\bibitem{Morrison:1994fr}
D.~R. Morrison and M.~Ronen~Plesser, ``{Summing the instantons: quantum
  cohomology and mirror symmetry in toric varieties},'' {\em Nucl. Phys.}
  {\bfseries B440} (1995) 279--354,
\href{http://arxiv.org/abs/hep-th/9412236}{{\ttfamily arXiv:hep-th/9412236}}.

\bibitem{Aspinwall:1993rj}
P.~S. Aspinwall, B.~R. Greene, and D.~R. Morrison, ``{The monomial divisor
  mirror map},'' {\em Internat. Math. Res. Notices} no.~12, (1993) 319--337,
\href{http://arxiv.org/abs/alg-geom/9309007}{{\ttfamily
  arXiv:alg-geom/9309007}}.

\bibitem{Cox:2000vi}
D.~A. Cox and S.~Katz, {\em {Mirror symmetry and algebraic geometry}}.
\newblock Amer. Math. Soc., 1999.
\newblock Providence, USA: AMS (2000) 469 p.

\bibitem{Kreuzer:2010ph}
M.~Kreuzer, J.~McOrist, I.~V. Melnikov, and M.~Plesser, ``{(0,2) deformations
  of linear sigma models},''
  \href{http://dx.doi.org/10.1007/JHEP07(2011)044}{{\em JHEP} {\bfseries 1107}
  (2011) 044}, \href{http://arxiv.org/abs/1001.2104}{{\ttfamily arXiv:1001.2104
  [hep-th]}}.

\bibitem{Melnikov:2010sa}
I.~V. Melnikov and M.~R. Plesser, ``{A (0,2) mirror map},''
  \href{http://dx.doi.org/10.1007/JHEP02(2011)001}{{\em JHEP} {\bfseries 1102}
  (2011) 001},
\href{http://arxiv.org/abs/1003.1303}{{\ttfamily arXiv:1003.1303 [hep-th]}}.

\bibitem{Berglund:1995gd}
P.~Berglund, S.~H. Katz, and A.~Klemm, ``{Mirror symmetry and the moduli space
  for generic hypersurfaces in toric varieties},''
  \href{http://dx.doi.org/10.1016/0550-3213(95)00434-2}{{\em Nucl. Phys. B}
  {\bfseries 456} (1995) 153--204},
  \href{http://arxiv.org/abs/hep-th/9506091}{{\ttfamily arXiv:hep-th/9506091}}.

\bibitem{Anderson:2015iia}
L.~B. Anderson, F.~Apruzzi, X.~Gao, J.~Gray, and S.-J. Lee, ``{A new
  construction of Calabi--Yau manifolds: Generalized CICYs},''
  \href{http://dx.doi.org/10.1016/j.nuclphysb.2016.03.016}{{\em Nucl. Phys.}
  {\bfseries B906} (2016) 441--496},
  \href{http://arxiv.org/abs/1507.03235}{{\ttfamily arXiv:1507.03235
  [hep-th]}}.

\bibitem{Guffin:2008kt}
J.~Guffin and E.~Sharpe, ``{A-twisted Landau-Ginzburg models},''
\href{http://arxiv.org/abs/0801.3836}{{\ttfamily arXiv:0801.3836 [hep-th]}}.

\bibitem{Jarvis:2014}
A.~Francis, T.~Jarvis, and N.~Priddis, ``A brief survey of {FJRW} theory,'' in
  {\em Primitive forms and related subjects --- Kavli IPMU 2014}, K.~Hori,
  C.~Li, S.~Li, and K.~Saito, eds., vol.~83 of {\em Adv. Stud. Pure Math.},
  pp.~19--53.
\newblock 2019.

\bibitem{Li:2019}
S.~Li and H.~Wen, ``On the {$L^2$}-{Hodge} theory of {Landau-Ginzburg
  }models,''
  \href{http://dx.doi.org/https://doi.org/10.48550/arXiv.1903.02713}{{\em Adv.
  Math.} {\bfseries 396} (2019) }. \url{https://arxiv.org/abs/1903.02713}.

\bibitem{Bertolini:2013xga}
M.~Bertolini, I.~V. Melnikov, and M.~R. Plesser, ``{Hybrid conformal field
  theories},''
\href{http://arxiv.org/abs/1307.7063}{{\ttfamily arXiv:1307.7063 [hep-th]}}.

\bibitem{Erkinger:2022sqs}
D.~Erkinger and J.~Knapp, ``{On genus-0 invariants of Calabi-Yau hybrid
  models},'' \href{http://arxiv.org/abs/2210.01226}{{\ttfamily arXiv:2210.01226
  [hep-th]}}.

\bibitem{Deligne:1999qp}
P.~Deligne, P.~Etingof, D.~S. Freed, L.~C. Jeffrey, D.~Kazhdan, J.~W. Morgan,
  D.~R. Morrison, and E.~Witten, eds., {\em {Quantum fields and strings: A
  course for mathematicians. Vol. 1, 2}}.
\newblock Amer. Math. Soc., 1999.

\bibitem{West:1990tg}
P.~West, {\em Introduction to supersymmetry and supergravity}.
\newblock World Scientific, Singapore, 1990.

\bibitem{Wess:1992cp}
J.~Wess and J.~Bagger, {\em {Supersymmetry and supergravity}}.
\newblock Princeton University Press, 1992.

\bibitem{Bertolini:2021hal}
M.~Bertolini, I.~V. Melnikov, and M.~R. Plesser, ``{Fixed points of (0,2)
  Landau-Ginzburg renormalization group flows and the chiral algebra},''
  \href{http://dx.doi.org/10.1007/JHEP09(2022)230}{{\em JHEP} {\bfseries 09}
  (2022) 230}, \href{http://arxiv.org/abs/2106.00105}{{\ttfamily
  arXiv:2106.00105 [hep-th]}}.

\bibitem{Beasley:2004ys}
C.~Beasley and E.~Witten, ``{New instanton effects in supersymmetric QCD},''
  \href{http://dx.doi.org/10.1088/1126-6708/2005/01/056}{{\em JHEP} {\bfseries
  0501} (2005) 056}, \href{http://arxiv.org/abs/hep-th/0409149}{{\ttfamily
  arXiv:hep-th/0409149 [hep-th]}}.

\bibitem{Green:2010da}
D.~Green, Z.~Komargodski, N.~Seiberg, Y.~Tachikawa, and B.~Wecht, ``{Exactly
  marginal deformations and global symmetries},''
  \href{http://dx.doi.org/10.1007/JHEP06(2010)106}{{\em JHEP} {\bfseries 1006}
  (2010) 106}, \href{http://arxiv.org/abs/1005.3546}{{\ttfamily arXiv:1005.3546
  [hep-th]}}.

\bibitem{Gomis:2016sab}
J.~Gomis, Z.~Komargodski, H.~Ooguri, N.~Seiberg, and Y.~Wang, ``{Shortening
  Anomalies in Supersymmetric Theories},''
  \href{http://dx.doi.org/10.1007/JHEP01(2017)067}{{\em JHEP} {\bfseries 01}
  (2017) 067},
\href{http://arxiv.org/abs/1611.03101}{{\ttfamily arXiv:1611.03101 [hep-th]}}.

\bibitem{Bertolini:2014ela}
M.~Bertolini, I.~V. Melnikov, and M.~R. Plesser, ``{Accidents in (0,2)
  Landau-Ginzburg theories},''
  \href{http://dx.doi.org/10.1007/JHEP12(2014)157}{{\em JHEP} {\bfseries 12}
  (2014) 157},
\href{http://arxiv.org/abs/1405.4266}{{\ttfamily arXiv:1405.4266 [hep-th]}}.

\bibitem{Gomis:2015yaa}
J.~Gomis, P.-S. Hsin, Z.~Komargodski, A.~Schwimmer, N.~Seiberg, and S.~Theisen,
  ``{Anomalies, Conformal Manifolds, and Spheres},''
  \href{http://dx.doi.org/10.1007/JHEP03(2016)022}{{\em JHEP} {\bfseries 03}
  (2016) 022},
\href{http://arxiv.org/abs/1509.08511}{{\ttfamily arXiv:1509.08511 [hep-th]}}.

\bibitem{Angella:2014ca}
D.~Angella, \href{http://dx.doi.org/10.1007/978-3-319-02441-7}{{\em
  Cohomological aspects in complex non-{K}\"ahler geometry}}, vol.~2095 of {\em
  Lecture Notes in Mathematics}.
\newblock Springer, Cham, 2014.
\newblock \url{https://doi.org/10.1007/978-3-319-02441-7}.

\bibitem{Witten:1993jg}
E.~Witten, ``{On the Landau-Ginzburg description of N=2 minimal models},'' {\em
  Int. J. Mod. Phys.} {\bfseries A9} (1994) 4783--4800,
\href{http://arxiv.org/abs/hep-th/9304026}{{\ttfamily arXiv:hep-th/9304026}}.

\bibitem{Silverstein:1995re}
E.~Silverstein and E.~Witten, ``{Criteria for conformal invariance of (0,2)
  models},'' {\em Nucl. Phys.} {\bfseries B444} (1995) 161--190,
\href{http://arxiv.org/abs/hep-th/9503212}{{\ttfamily arXiv:hep-th/9503212}}.

\bibitem{Witten:2005px}
E.~Witten, ``Two-dimensional models with (0,2) supersymmetry: Perturbative
  aspects,''
\href{http://arxiv.org/abs/hep-th/0504078}{{\ttfamily hep-th/0504078}}.

\bibitem{Malikov:1998dw}
F.~Malikov, V.~Schechtman, and A.~Vaintrob, ``{Chiral de Rham complex},''
  \href{http://dx.doi.org/10.1007/s002200050653}{{\em Commun. Math. Phys.}
  {\bfseries 204} (1999) 439--473},
  \href{http://arxiv.org/abs/math/9803041}{{\ttfamily arXiv:math/9803041}}.

\bibitem{Dedushenko:2015opz}
M.~Dedushenko, ``{Chiral algebras in Landau-Ginzburg models},''
\href{http://arxiv.org/abs/1511.04372}{{\ttfamily arXiv:1511.04372 [hep-th]}}.

\bibitem{Bertolini:2018now}
M.~Bertolini and M.~Romo, ``{Aspects of (2,2) and (0,2) hybrid models},''
\href{http://arxiv.org/abs/1801.04100}{{\ttfamily arXiv:1801.04100 [hep-th]}}.

\bibitem{Adams:2005tc}
A.~Adams, J.~Distler, and M.~Ernebjerg, ``Topological heterotic rings,'' {\em
  Adv. Theor. Math. Phys.} {\bfseries 10} (2006) 657--682,
\href{http://arxiv.org/abs/hep-th/0506263}{{\ttfamily hep-th/0506263}}.

\bibitem{Kreuzer:2000xy}
M.~Kreuzer and H.~Skarke, ``{Complete classification of reflexive polyhedra in
  four dimensions},'' {\em Adv. Theor. Math. Phys.} {\bfseries 4} (2002)
  1209--1230,
\href{http://arxiv.org/abs/hep-th/0002240}{{\ttfamily arXiv:hep-th/0002240}}.

\bibitem{Aspinwall:2010ve}
P.~S. Aspinwall, I.~V. Melnikov, and M.~R. Plesser, ``{(0,2) Elephants},'' {\em
  JHEP} {\bfseries 1201} (2012) 060,
  \href{http://arxiv.org/abs/1008.2156}{{\ttfamily arXiv:1008.2156 [hep-th]}}.

\bibitem{Cox:2011tv}
D.~Cox, J.~Little, and H.~Schenck, {\em Toric varieties}, vol.~124 of {\em
  Graduate Studies in Mathematics}.
\newblock AMS, 2011.

\bibitem{Mavlyutov:2003fa}
A.~R. Mavlyutov, ``{Deformations of Calabi-Yau hypersurfaces arising from
  deformations of toric varieties},''
  \href{http://dx.doi.org/10.1007/s00222-004-0362-7}{{\em Invent. Math.}
  {\bfseries 157} (2004) 621},
  \href{http://arxiv.org/abs/math/0309239}{{\ttfamily arXiv:math/0309239}}.

\bibitem{Demailly:1994}
J.-P. Demailly, T.~Peternell, and M.~Schneider, ``Compact complex manifolds
  with numerically effecitve tangent bundles,'' {\em J. Alg. Geom.} {\bfseries
  3} (1994) 295--345.
  \url{https://www-fourier.ujf-grenoble.fr/~demailly/manuscripts/dps1.pdf}.

\bibitem{Atiyah:1957xx}
M.~F. Atiyah, ``Complex analytic connections in fibre bundles,'' {\em Trans.
  AMS} {\bfseries 85} no.~1, (1957) 181--207.

\bibitem{Chen:2019wub}
Q.~Chen, F.~Janda, and Y.~Ruan, ``{The logarithmic gauged linear sigma
  model},'' \href{http://dx.doi.org/10.1007/s00222-021-01044-2}{{\em Invent.
  Math.} {\bfseries 225} no.~3, (2021) 1077--1154},
  \href{http://arxiv.org/abs/1906.04345}{{\ttfamily arXiv:1906.04345
  [math.AG]}}.

\end{thebibliography}\endgroup

\end{document}